\documentclass[]{article}

\usepackage{arxiv}
\usepackage[utf8]{inputenc} 
\usepackage[T1]{fontenc}    
\usepackage{hyperref}       
\usepackage{url}            
\usepackage{booktabs}       
\usepackage{amsfonts}       
\usepackage{nicefrac}       
\usepackage{microtype}      
\usepackage{lipsum}		
\usepackage{graphicx}
\usepackage{natbib}
\usepackage{doi}
\usepackage{units}
\usepackage{amsmath}
\usepackage{tabulary}

\newcommand{\eq}[1]{Equation~#1} 
\newcommand{\fig}[1]{Figure~#1} 
\newcommand{\sect}[1]{Section~#1}
\newcommand{\Sect}[1]{Section~#1}

\newcommand{\tab}[1]{Table~#1}
\newcommand{\app}[1]{Appendix~#1}

\newcommand{\Dref}{\widetilde{D}}
\newcommand{\flux}{\mathcal{F}}
\newcommand{\Fvolume}{\mathcal{V}}

\newcommand{\refBase}{\tilde{\phi}}
\newcommand{\bandpass}{\mathcal{B}}
\newcommand{\contravec}{\hat{\mathbf{e}}}
\newcommand{\covec}{\hat{\mathbf{e}}\T}

\newcommand{\ba}{\mathbf{a}}
\DeclareMathOperator{\spectrum}{\mathcal{S}}
\newcommand{\dr}{\Delta r}
\newcommand{\legendre}[1]{\tilde{\phi}_{#1}}

\newcommand{\coordTrans}{\mathbf{C}}

\newcommand{\bx}{\mathbf{x}}

\newcommand{\bxnAna}{\bx^{\rm{a},(n)}}
\newcommand{\by}{\mathbf{y}}
\newcommand{\xdg}{x^{\rm{DG}}}
\newcommand{\xtrue}{x^{\rm{truth}}}
\newcommand{\Nobs}{N_{\rm{obs}}}
\newcommand{\bxdg}{\bx^{\rm{DG}}}
\newcommand{\bxp}{\bx^{\rm{NM}}}

\newcommand{\bBdg}{\bB^{\rm{DG}}}

\newcommand{\bxMean}{\overline{\bx}}
\newcommand{\bxMeanFor}{\overline{\bx}^{\rm{f}}}
\newcommand{\bxMeanAna}{\overline{\bx}^{\rm{a}}}
\newcommand\br{\mathbf{r}}
\newcommand{\nextsimdg}{neXtSIM$_{DG}$}
\newcommand{\nextsim}{neXtSIM}
\newcommand{\pd}[2]{\frac{\partial #1}{\partial #2}}
\newcommand{\projection}{\mathcal{P}}
\newcommand{\fourier}{\mathcal{F}}

\DeclareMathOperator{\sign}{\mathrm{sign}}
\DeclareMathOperator{\trans}{\mathcal{T}}
\DeclareMathOperator{\mirror}{\mathcal{S}}
\newcommand{\inno}{\mathbf{d}}
\newcommand{\expectation}[1]{\mathbb{E}[#1]}
\DeclareMathOperator{\bigo}{\mathcal{O}}
\newcommand\bB{\mathbf{B}}
\newcommand\bR{\mathbf{R}}
\newcommand\bI{\mathbf{I}}
\newcommand\bH{\mathbf{H}}
\newcommand\bK{\mathbf{K}}
\newcommand\bL{\mathbf{L}}
\newcommand\R{\mathbb{R}}
\newcommand{\T}{^\top}
\newcommand{\bchi}{\mathbf{\chi}}

\title{Tailoring data assimilation to discontinuous Galerkin models}

\author{ \href{https://orcid.org/0000-0001-5076-5421}{\includegraphics[scale=0.06]{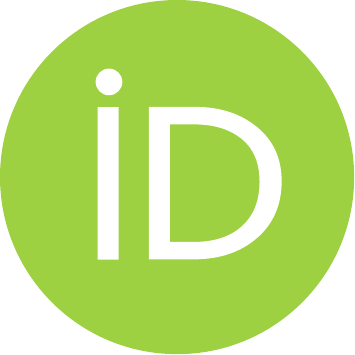}\hspace{1mm}Ivo Pasmans} \\
    University of Reading, Reading, UK \\
	\texttt{i.c.pasmans@reading.ac.uk} \\
	\And
	\href{http://orcid.org/0000-0002-2319-6937}{\includegraphics[scale=0.06]{orcid.pdf}\hspace{1mm}Yumeng Chen} \\
	University of Reading, Reading, UK\\
	\And 
	\href{https://orcid.org/0000-0003-0722-5600}{\includegraphics[scale=0.06]{orcid.pdf}\hspace{1mm}Alberto Carrassi} \\
	Department of Physics ``Augusto Righi'', Universit\`a di Bologna, Bologna, Italy \\
	University of Reading, Reading, UK\\
	\And
    Chris K.R.T. Jones \\
	University of North Carolina, Chapel Hill NC, USA
}

\begin{document}

\maketitle
\begin{abstract}
During the last few years discontinuous Galerkin (DG) methods have received increased interest from the geophysical community. In these methods the solution in each grid cell is approximated as a linear combination of basis functions. Ensemble data assimilation (EnDA) aims to approximate the true state by combining model outputs with observations using error statistics estimated from an ensemble of model runs. Ensemble data assimilation in geophysical models faces several well-documented issues. In this work we exploit the expansion of the solution in DG basis functions to address some of these issues. Specifically, it is investigated whether a DA-DG combination 1) mitigates the need for observation thinning, 2) reduces errors in the field's gradients, 3) can be used to set up scale-dependent localisation. Numerical experiments are carried out using stochastically generated ensembles of model states, with different noise properties, and with Legendre polynomials as basis functions. It is found that strong reduction in the analysis error is achieved by using DA-DG and that the benefit increases with increasing DG order. This is especially the case when small scales dominate the background error. The DA improvement in the 1st derivative is on the other hand marginal. We think this to be a counter-effect of the power of DG to fit the observations closely, which can deteriorate the estimates of the derivatives. Applying optimal localisation to the different polynomial orders, thus exploiting their different spatial length, is beneficial: it results in a covariance matrix closer to the true covariance than the matrix obtained using traditional optimal localisation in state-space. 

\end{abstract}%

\section{Introduction}
{\label{sec:intro}}

The discontinuous Galerkin (DG) method is a numerical method to the solution of a system of partial differential equations \citep{cockburn_discontinuous_2003}. It does this by dividing the model domain into grid cells and assuming that in each of the grid cells the solution can be approximated by a finite linear combination of functions, usually polynomials. On the grid cells' edges the solution is not uniquely defined and may contain discontinuities, hence the method's name. Discontinuous Galerkin methods combine several advantageous features from other numerical methods like finite-difference, finite-volume and finite-element: DG schemes conserve tracer volumes, mass and momentum (though not necessarily energy); flux calculations involve at most two grid cells, thus allowing for better scalability than other methods; likewise finite volume methods DG can be applied to unstructured meshes; and as discontinuities are allowed at the grid cell edge it is relatively straightforward to merge/divide cells ({\it h-adaptivity}) and/or vary the number of basis functions ({\it p-adaptivity}) on a cell-by-cell basis \citep{marras_review_2016-1}. Furthermore, for sufficiently smooth solutions, increasing the polynomial order turns out to be a computationally more efficient way to reach a desired accuracy than reducing the grid cell size \citep{vos_h_2010}. 

The potential benefits of DG solvers have not escaped the attention of the geophysical community. Over the last few years, several numerical geophysical prediction (research) systems have been developed/are under development that use (partial) DG methods in the computational core. Examples of such efforts are the coastal ocean model Thetis \citep{karna_thetis_2018}, the U.S. Navy/Air Force NUMA atmospheric model \citep{giraldo_development_2011}, the soil water flow model {DORiE} \citep{riedel_dorie_2020}, or the sea-ice model {\nextsimdg} being developed as part of the Scale-Aware Sea Ice Project (SASIP; \url{https://sasip-climate.github.io/}). The latter development is the direct motivation for the work laid out in this paper. {\nextsimdg} uses the Maxwell-elasto-brittle rheology and thermodynamics from its ancestor, the {\nextsim} sea-ice model \citep{bouillon_presentation_2015,rampal_nextsim_2016, samake_parallel_2017}, but replaces the original finite element numeric on an adaptive Lagrangian mesh with a DG numeric on an Eulerian mesh.  This eases the embedding in a global climate system, keeps accurate representation of sharp gradients in the physical fields and facilitates efficient parallelisation of the model to achieve superior computational performance. 

In the geosciences and climate enterprise it is common practice to periodically correct model state towards the truth using observations in a process called data assimilation (DA). A wide range of DA algorithms is available for this purpose. See e.g. \citet{navon_data_2009,bannister_review_2017, carrassi_data_2018,evensen_data_2022} for an overview. In particular, DA is used in sea ice models: Environment Canada \citep{buehner_new_2013}, Artic Cap Nowcast/Forecast System (ACNFS) \citep{hebert_short-term_2015} and The Multivariate Ocean Variational Estimation/Meteorological Research Institute Community Ocean Model (MOVE/MRI.COM) \citep{toyoda_data_2016} use 3DVar, MITgcm has been paired with a Local Error Subspace Transform Kalman filter \citep{mu_arctic-wide_2018}, while for TOPAZ4 \citep{sakov_topaz4_2012,xie_impact_2018}, NEMO-LIM2 \citep{massonnet_model_2013} and {\nextsim} \citep{sampson_ensemble_2021-1, cheng_arctic_2023, richter_nextsim-dg_2023} various versions of the ensemble Kalman filter (EnKF) \citep{evensen_data_2022} have been used. It is planned that DA capabilities will also be added to \nextsimdg. Ensemble Kalman filters are model agnostic and can in principle directly be used with DG-based models such as \nextsimdg. At the observation times, the EnKF will update the model fields propagated and expressed via the DG method, but it will be blind to the underlying DG structure. By allowing the EnKF to act directly at the level of the DG decomposition of the physical fields, we believe that we can improve the DA performance, stability and efficiency. This is the overarching objective of this study. 

We hypothesise that it is possible to exploit the structure of a DG solver directly to get better DA performance at little additional computational cost. In particular, we will investigate whether it is possible to 1) exploit the fact that DG basis functions resolve the solution at a subcell level to assimilate multiple observations per grid cell; 2) correct not only the field itself but also its derivatives and 3) profit from the separation of the solution in basis functions inherent to DG to construct a scale-dependent localisation scheme. We plan to investigate these hypotheses using a synthetic, univariate 1D-model. I.e. a model in which the states are realisations of a statistical distribution and in which a dynamical forward model is absent. 

The first and second goal of this study are motivated by the density of observations bestowed by the satellites. The resolution for the latest generation of satellite missions is within $\bigo(10^1)-\bigo(10^3) \,\unit{m}$ \citep{emery_arctic_1994,drue_high-resolution_2004,drusch_sentinel-2_2012,gong_finer_2013,gourmelen_cryosat-2_2018}, which is already smaller than typical grid sizes in numerical models ($\bigo(10^{3}) \,\unit{m}$). In standard DA methods and in the absence of representativeness errors or other spatial correlation between observational errors, the number of degrees of freedom that can be fit (controlled) using data for each physical field is bounded to the number of grid cells. Consequently, if one attempts to assimilate on average more than $1$ observation per grid cell it is not possible for DA to fit all observations simultaneously. In has been shown that attempting to assimilate more than 1 observation per grid cell under such circumstances will render little improvement \citep{liu_interaction_2002}. In that situation DA will not be able to incorporate all the information contained in, e.g. the huge volume of high-resolution satellite observations. This results not only in an underutilisation of the information in the observations, but it also needlessly consumes computational resources during the DA process. In the worst cases, it might even degrade the solution as the portion of the observational signal that cannot be resolved by the model gives rise to {\it representation errors} \citep{janjic_representation_2018}, a ``thorn in the flesh'' for DA. These issues are usually coped and mitigated by data thinning, the process of selecting only a portion of available data, and by {\it superobbing}. The latter is the process of averaging all observations from the same observation platform over one (or more) grid cell(s) into a single observation \citep{oke_bluelink_2008}. Although very popular, as it represents a concrete pragmatic option, superobbing is not ideal as information in the small scales is lost in the process. In DG models the solution is (partially) resolved at the subcell scale. We therefore argue that less superobbing and data thinning are required when doing DA with DG models. An improved capability to assimilate densely spaced data should then open up to the possibility for DA to correct smaller scales, thus improving  1st- (the gradient), or even higher-order derivatives, of the model fields. 

Our third research question, DG-based scale-dependent localisation, tackles a central concern for ensemble-based DA methods in high dimensions. Ensemble-based DA algorithms use an ensemble of model runs to estimate $\bB$, the covariance of the errors in the model state. Due to the finite-size of the ensemble, the ensemble-estimate is plagued by sampling errors \citep{houtekamer_data_1998}. If the absolute error correlation between the elements that make up the state vector is small (as it is usually the case in physical systems where correlation decays exponentially with distance; \citet[see e.g.,][]{carrassi_data_2018}) sampling errors can dominate the true correlation. The process to suppress this sampling error is referred to as localisation. Two categories of localisation scheme are in common use \citep{carrassi_data_2018}: domain localisation and covariance localisation. Though computationally more efficient, domain localisation is not suitable for observations that are non-local or exhibit spatially correlated observation errors. As such conditions are common when dealing with dense satellite observations, the focus of this paper will be on covariance localisation in which localisation is applied directly to the background error covariance. In its most prevalent form, covariance localisation is applied in state space: it is assumed that the true error correlations tend to zero over long distances and this behaviour is imposed on the ensemble covariance by tapering it off to zero with increasing distance \citep{hamill_distance-dependent_2001-3,ehrendorfer_review_2007}. The distance at which the sampling noise dominates the true error covariance will depend on the scales in the true covariance: the larger the scale in the covariance, the larger the tapering, or localisation, distance should be. However, distance-based localisation schemes in the state space do not offer this flexibility. Alternatively, the covariance can be localised in Fourier space \citep{buehner_spectral_2007}. This makes localisation scale dependent, but this dependence cannot be varied with position in the model domain. Localisation in wavelet-space combines the best parts of the preceding two approaches \citep{tangborn_wavelet_2004,deckmyn_wavelet_2005-1,varella_diagnostic_2011,berre_modelling_2015,chabot_diagnosis_2017}. As each wavelet represents a limited range of scales for a limited region, the localisation can be adjusted based on both scale and position. Multiple families of wavelets, e.g. orthogonal Meyer and spherical harmonics, have been used. In previous applications the use of wavelets required transformation of model output. As current generation of geophysical models are formulated in either space or spectral space, wavelet-based localisation in an ensemble DA requires transformation of ensemble members to and from wavelet space. We will show in this paper that DG basis functions can act as model-intrinsic
wavelet basis thus eliminating the need for wavelet transformations. The third aim of this paper is to investigate if this model-intrinsic wavelet space can be used for localisation. 

The paper is organized as follows. \Sect{\ref{sec:dg}} describes the general setup of a DG model like \nextsimdg. This is followed in \sect{\ref{sec:da}} by a general description of DA scheme used in this study. \Sect{\ref{sec:superob}} contains the setup and results of several univariate 1D twin-experiments using synthetic truths and ensembles. With these we look at the benefits of assimilating multiple observations per cell on the model field and on its derivatives. \Sect{\ref{sec:localisation}} describes the framework of our proposed scale-dependent localisation scheme and numerical experiments.  Finally, \sect{\ref{sec:conclusion}}, summarises our findings, draws the conclusion and the forward perspectives of our study. Thereby paving the way to the application of DG-informed DA in more realistic settings. 

\section{Discontinuous Galerkin Methods}
{\label{sec:dg}}

In the last decades the finite-volume method has become one of the most popular numerical methods to solve partial differential equations in geophysical models \citep{shchepetkin_regional_2005, chen_unstructured_2006,  kuhnlein_fvm_2019, harris_scientific_2021-1, adcroft_mitgcm_2022} due to the fact that they inherently obey mass and tracer conservation laws. In the finite-volume method the unknowns are assumed to be constant within a model grid cell and their changes in time are determined by the flux through the grid cell edges. Equivalently, one can say that the solution in the grid cell is approximated by a multiple of a $0$th-order polynomial. This suggests a straightforward extension to a scheme in which the solution in each grid cell is approximated as a linear combination of higher-order polynomials. This gave rise to discontinuous Galerkin (DG) methods \citep{hesthaven_nodal_2007}. A short overview of the general structure of univariate DG models is presented in this section. Hereafter we will use italic symbols to indicate quantities that are piecewise continuous functions of space while we will use bold symbols for quantities that can be represented by finite arrays and therefore can be stored in computer memory. 

Suppose that the exact solution at position $r$, time $t$, to a system of partial differential equations is given as $\xtrue(r,t)$. Then the DG approximation to the scalar function $\xdg(r,t)$ is given by 
\begin{equation}
     \xdg(r,t) = \sum_{l=0}^{L-1}\sum_{m=1}^{M} \bxdg_{lm}(t) \phi_{lm}(r)
    \label{eq:expansion}
\end{equation}
with $M$ the number of grid cells, $L$ the number of basis polynomials used in the DG approximation and $\bxdg(t) \in \R^{L \times M}$ a matrix. The  (scalar) elements of this matrix, the DG coefficients $\bxdg_{lm}$, depend only on time while the spatial part of the DG solution is encoded in the basis functions
\begin{equation}
    \phi_{lm}(r) =
    \begin{cases}
        \sqrt{|\det D\Psi_{m}^{-1}(r)|}\refBase_{l}\circ \Psi^{-1}_{m}(r), & \text{if } r \in D_{m}\\
        0 & \text{otherwise}
    \end{cases}
    \label{eq:support}
\end{equation}
 where $\Psi_{m}:\Dref \to D_{m}$ is a coordinate transform from a reference domain $\Dref$ to the $m$th grid cell $D_{m}$ with $\det D\Psi_{m}(r)$ the determinant of its Jacobian, $D$ being the model domain, and $\refBase_k = \{\refBase_{l} : 0 \leq l<L-1\}_{l}$ is a set of reference functions defined on $\Dref$. Specifically, each basis function is defined locally within a grid cell and irrelevant outside of it. The set $\phi = \{\phi_{lm}: 0\leq l < L_{k},\, 1\leq m \leq M\}$ endowed with the $L^2$-inner product defines a Hilbert space of functions denoted here by $V$. 

Several function families provide suitable candidates for the basis functions $\refBase_k$. On a top level, bases can be divided between {\it nodal} and {\it modal} bases \citep[][section 3.1]{hesthaven_nodal_2007}. For nodal basis functions the DG coefficients are equal to the solution evaluated at nodal (grid) points, i.e. $\bxp_{lm}(t) = \bxp(r_{lm},t)$
if $r_{lm} \in D_{m}$. An 1D example of such a nodal basis is the family of Lagrange polynomials with $\refBase_{l}$ defined as the unique $(L-1)$th-order polynomial that satisfies $\refBase_{l}(r_{l'})=\delta_{ll'}$ for all $0 \leq l < L$ and $\delta$ the Kronecker delta. In contrast, DG coefficients for a modal basis cannot be directly found by evaluating the function, but require projecting the solution onto the modal basis functions. One such example of a modal basis in 1D, and the one that will be used in this paper, is the  Legendre basis. In this case $\Dref=[-1,1]$, $\Psi_{m}(r)=\frac{r_{m+1}+r_{m}}{2}+r\frac{r_{m+1}-r_{m}}{2}$ with $r_{m}$ the left side of the $m$th grid cell and $\refBase_{l}(r)=P_{l}(r)$ the $l$th-order Legendre polynomial. In this case, the DG coefficient $\bxdg_{lm}$ can be found by projecting $\xdg$ on the basis polynomial:
\begin{eqnarray}
    && \int_{D} \phi_{lm}(r) \xdg(r,t) \,\mathrm{d}r
    = \sum_{l'=0}^{L-1} \sum_{m'=1}^{M} \bxdg_{l'm'}(t) \int_{D} \phi_{lm}(r)\phi_{l'm'}(r)  \,\mathrm{d}r \nonumber \\
    &&= \sum_{l'=0}^{L-1} \bxdg_{l'm}(t) \int_{r_{m}}^{r_{m+1}} \phi_{lm}(r)\phi_{l'm}(r)  \,\mathrm{d}r 
    = \sum_{l'=0}^{L-1} \bxdg_{l'm} \int_{-1}^{1} P_{l}(r)P_{l'}(r) \,\mathrm{d}r = \bxdg_{lm}
     \label{eq:model_coef}
\end{eqnarray}
where $\xdg$ has been expanded using \eq{\ref{eq:expansion}} and the orthogonality of the Legendre polynomials, $\int_{-1}^{1} P_{l}(r) P_{l'}(r) \, \mathrm{d}r \sim \delta_{ll'}$, has been used. 
 
 In its most general form, a nonlinear geophysical model like {\nextsimdg} can be written as 
\begin{equation}
    0 = \mathcal{R}(\xtrue,t) 
    \overset{def}{=}  \pd{\mathcal{A}(\xtrue,t)}{t} -
    \left[  
        -\nabla \cdot \mathcal{F}(\xtrue,t)+\mathcal{V}(\xtrue,t) 
    \right]
    \label{eq:general}
\end{equation}
where $\mathcal{F}$ is the flux, and $\mathcal{V}$ the volume force. $\mathcal{A}$ can be a  nonlinear operator, but often is equal to the identity operator, i.e. $\mathcal{A}(x,t)=x$ and $\mathcal{R}$ is the residual operator. In the Bubnov-Galerkin method \citep{bellman_chapter_1970} an approximation $x \in V$ is sought such that the residual in \eq{\ref{eq:general}} is orthogonal to $V$, i.e. for all $0 < l \leq L-1$, $1 < m \leq M$ it must hold that

\begin{eqnarray}
    0 &=& \int_{D} \phi_{lm}  \mathcal{R}(x,t)  \,\mathrm{d}r  
    = \pd{}{t}\int_{D} \phi_{lm} \mathcal{A}(x,t) \,\mathrm{d}r
   +\int_{D} \phi_{lm} \nabla \cdot \flux(x,t) \,\mathrm{d}r
   - \int_{D} \phi_{lm} \Fvolume(x,t) \,\mathrm{d}r  \nonumber \\ 
    &=& \pd{}{t}\int_{D_{m}} \phi_{lm} \mathcal{A}(x,t) \,\mathrm{d}r
   + \int_{\partial D_{m}} \phi_{lm} \flux(x,t)\cdot dS 
   - \int_{D_{m}} (\nabla \phi_{lm}) \cdot \flux(x,t) \,\mathrm{d}r 
   - \int_{D_{m}} \phi_{lm} \Fvolume(x,t) \,\mathrm{d}r
   \label{eq:general_dg}  \nonumber \\
   &\approx& 
   \pd{}{t}\int_{D_{m}} \phi_{lm} \mathcal{A}(x,t) \,\mathrm{d}r
   + \int_{\partial D_{m}} \phi_{lm} \flux^{*}(x,t)\cdot \mathrm{d}S \nonumber \\
   &-& \int_{D_{m}} (\nabla \phi_{lm}) \cdot \flux(x,t) \,\mathrm{d}r
   - \int_{D_{m}} \phi_{lm} \Fvolume(x,t)\,\mathrm{d}r
\end{eqnarray}
where $\partial D_{m}$ is the boundary of the $m$th cell and $\mathrm{d}S$ the boundary area element and $\flux^{*}$ is the numerical flux that will be discussed below. In the second line integration-by-parts has been used in combination with the information on the local support of $\phi_{lm}$ given in \eq{\ref{eq:support}}.

The flux $\mathcal{F}(x,t)$ is not uniquely defined on the grid cell boundaries since $\phi_{lm}$ exhibit discontinuities here. The fundamental {\it ansatz} made in DG methods is that this flux can be replaced by a numerical flux $\flux^{*}=\flux^{*}(x_{-},x_{+},t)$. Here $x_{-}$ is the approximation of $x$ on one side of the cell boundary and $x_{+}$ on the other side. In case the cell boundary is part of the domain's outer boundary, $x_{+}$ is provided by the boundary conditions. The numerical flux is problem specific and has to be chosen with care as the stability of the DG scheme depends on it \citep[][section 4.4]{hesthaven_nodal_2007}. 

After integration, \eq{\ref{eq:general_dg}} can be represented as a  system of ordinary differential equations for $\bxdg_{lm}$ with  $0\leq l < L$, $1\leq m \leq M$
\begin{equation}
    \frac{\mathrm{d}}{\mathrm{d}t} \tilde{\mathcal{A}}_{lm}(\bxdg,t) = 
    -\tilde{\mathcal{S}}_{lm}(\bxdg,t)+\tilde{\mathcal{M}}_{lm}(\bxdg,t)
    \label{eq:general_c}
\end{equation}
with $\bxdg$ a tensor having $\bxdg_{lm}$ as its elements and
\begin{eqnarray*}
    \tilde{\mathcal{A}}_{lm}(\bxdg,t) &=&  \int_{D} \phi_{lm}(r) 
    \mathcal{A}(\sum_{l} \sum_{m'} \bxdg_{l'm'}(t) \phi_{l'm'}(r),t) \,\mathrm{d}r \\
    \tilde{\mathcal{M}}_{lm}(\bxdg,t) &=& 
    \int_{D} \nabla\phi_{lm}(r,t) \cdot \mathcal{F}(\sum_{l'm'} \bxdg_{l'm'}(t) \phi_{l'm'}(r),t) \,\mathrm{d}r \\
    &+& \int_{D} \phi_{lm}(r) \mathcal{V}(\sum_{l'm'} \bxdg_{l'm'}(t) \phi_{l'm'}(r),t)  \,\mathrm{d}r \\
    \mathcal{S}_{lm}(\bxdg,t) &=& 
    \sum_{m' < n'} \int_{\partial D_{m'}  \cap \partial D_{n'}} 
    \flux^{*}\Big( \sum_{l'} \bxdg_{l'm'} \phi_{l'm'}(r,t),
    \sum_{l'} \bxdg_{l'n'} \phi_{l'n'}(r,t) \Big)  \,\mathrm{d}S\\
    &+&  \sum_{m'} \int_{\partial D_{m'}  \cap \partial D} 
    \flux^{*}\Big( \sum_{l'} \bxdg_{l'm'} \phi_{l'm'}(r,t),
    \sum_{l'} \bxdg_{l'm'} \phi_{l'm'}(\br,t) \Big) \,\mathrm{d}S
\end{eqnarray*}

The preceding discussion can easily be extended to multivariate models by expanding each model field separately.

\section{Data Assimilation with DG-based models}
{\label{sec:da}}

Data assimilation (DA) combines a prior guess of the model state, also known as the background, with observations in order to produce an improved estimate of the true state of the system. A plethora of DA methods have been developed over the past decades: though different in their details, all share some basic components. Among these are a list of observations and the specification of a probability distribution for the observational errors, the specification of the model state as a finite array of elements together with an estimate of the {\it apriori} error distribution of these elements, the background error, and an observation operator that takes the model state as input and outputs predictions for the observed quantities. 

We shall study the implications, advantages, drawbacks and ways of adaption of DA to DG models. As a prototype of a DA method, we will work with the Deterministic Ensemble-3DVar (D-E3DVar): an ensemble-variational method based on the Deterministic Ensemble Kalman Filter \citep{sakov_deterministic_2008}. In the version of D-E3DVar used here it is assumed that observational error statistics are Gaussian, static in time and that there is no correlation between the errors of different observations, while background errors are assumed to be Gaussian with a varying covariance estimated from an ensemble. Our choice for a method as D-E3DVar is motivated by the fact that it is an ensemble method and thus requires localisation. It is therefore amenable to study scale-aware localisations as we shall discuss in \sect{~\ref{sec:localisation}}. As variational method, the implementation of covariance localisation in the linear solver is straightforward, certainly when compared with an ensemble Kalman filter method. Furthermore, its results are easily extendable to ensemble of 4DVars systems that are currently at the forefront of DA development \citep{bonavita_evolution_2016, pasmans_ensemble_2019, ngodock_ensemble_2020, pasmans_ensemble_2020,zhu_4denvar-based_2022} and theoretically can deal with nonlinear observation operators. 

D-E3DVar DA updates the forecast state by minimising the cost-function
\begin{equation}
    J(\delta \bx) = \frac{1}{2} \delta \bx \T \bB^{-1} \delta \bx + \frac{1}{2}[\by-H(\bxMean^{f}+\delta \bx) ]\T \bR^{-1} [ \by-H(\bxMean^{f}+\delta \bx) ]
    \label{eq:cost}
\end{equation}
where $\bxMeanFor$ is the ensemble mean of the forecast states, $\delta \bx$ the DA correction (the {\it analysis increment}) to the forecast mean, $\by$ the vector containing the different observations, $H$ the (possibly nonlinear) observation operator mapping the model state into predictions for the observations and $\bR$ the observation error covariance matrix which we assume to be diagonal and constant in time. The background error covariance $\bB$ is estimated from an ensemble of $N$ independent model realizations as $\bB = \frac{1}{N-1} \sum_{n=1}^{N} \ba^{(n)} \otimes (\ba^{(n)})\T$, 
with $\ba^{(n)}=\bx^{(n)}- \bxMeanFor$ the forecast ensemble anomaly of ensemble member $n$. 

For the purpose of this study we conveniently consider point observations of the system's state vector, therefore $H$ is linear and we can write  
\begin{equation}
\by - H(\bxMeanFor+\delta \bx) = 
\by - \bH \bxMeanFor - \bH \delta \bx \overset{def}{=} \inno - \bH \delta \bx.
\end{equation}
The post-DA model state, i.e. the {\it analysis}, produced by minimising the cost function in \eq{\ref{eq:cost}} is given by 
\begin{equation}
\bxMeanAna= \bxMeanFor+\widehat{\delta \bx}  = \bxMeanFor + \bK \inno 
\label{eq:meanUpdate}
\end{equation}
with $\widehat{\delta \bx}$ the value for which $J$ attains its minimum. 
Similar to \citet{sakov_deterministic_2008} the members of the analysis ensemble are approximated as 
\begin{equation}
\bxnAna= \bxMeanAna + \ba^{(n)} 
-\frac{1}{2}\bK \bH \ba^{(n)}
\label{eq:ensUpdate}
\end{equation}
where $\bK= \bB \bH\T (\bH \bB \bH\T + \bR)^{-1}$ is the Kalman gain. 

The values $\bK \inno$ and $\bK \bH \ba^{(n)}$ are given as $\bB \bH\T \bR^{-\frac{1}{2}} \chi^{(0)}$ and
$\bB \bH\T \bR^{-\frac{1}{2}} \chi^{(n)}$ respectively where $\chi^{(n)}$ $0 \leq n \leq N$ are found by iteratively solving 
\begin{equation}
(\bR^{-\frac{1}{2}} \bH \bB \bH\T \bR^{-\frac{1}{2}} + \bI) \left[\bchi^{(0)}, \bchi^{(1)},\ldots,\bchi^{(N)}] =  [\bR^{-\frac{1}{2}} \inno, \bR^{-\frac{1}{2}} \bH \ba^{(1)},\ldots,\bR^{-\frac{1}{2}} \bH \ba^{(N)} \right]
\label{eq:rcg}
\end{equation}
using Reduced Conjugate Gradient method \citep{gurol_b-preconditioned_2014} and a block-diagonal Krylov algorithm \citep{auligne_ensemblevariational_2016,mercier_speeding_2019}.

With the DA scheme in hands, here the D-E3DVar
\eq{\ref{eq:cost}}--\ref{eq:ensUpdate}, one can then decide whether the DA correction should act on the model physical quantities at grid nodes $\bxp$, e.g. standard ``nodal approach'', or in the DG space, e.g. on the modal DG coefficients $\bxdg$. Assimilating in the DG or in the nodal space implies defining different observation operators. Observation operators for point observation at location $r$ are given in the second row of \tab{\ref{tab:interpolators}} together with observations of the 1st (3rd row) and 2nd derivative (4th row) at that location in a 1D model. For DG models evaluation of the expansion in \eq{\ref{eq:expansion}} provides a natural way to define these operators: the basis functions or their derivatives are evaluated at $r$ and multiplied with their respective DG coefficient. In the standard approach, model values at the left-side of cell $m$ are stored in $\bxp_{m}$ and linear interpolation is used to find values at arbitrary position $r$ (2nd row, 2nd column). The first (second) derivative of the model field are approximated in the grid cell centres (nodes) using a finite  difference method. Values of for the derivatives at $r$ are then obtained using linear interpolation of the field obtained by the finite difference method. The observational operator for the first (second) derivative thus obtained are given in the ``DG'', and  row ``0'' (``1'') of \tab{\ref{tab:interpolators}}.

\begin{table}[!h]
\centering
\normalsize\
\begin{tabulary}{1.0\textwidth}{lll}
\hline 
Derivative \#th & DG & Nodal \\
\hline

$0$ & $\sum_{l=0}^{L-1} \sum_{m=1}^{M} \bxdg_{lm} \phi_{lm}(r) = \bH^{\rm{DG}}_{r}$ 
& $\sum_{m=1}^{M} \bxp_{m} \chi_{m}(r) = \bH^{\rm{NM}}_{r}$  \\

$1$ & $\sum_{l=0}^{L-1} \sum_{m=1}^{M} \bxdg_{lm} \frac{\mathrm{d}\phi_{lm}}{\mathrm{d}r}(r)$ 
& $\sum_{m=1}^{M} \frac{\bxp_{m+1}-\bxp_{m}}{\dr} \chi_{m}(r-\frac{1}{2}\dr)$ 

\\
$2$ & $\sum_{l=0}^{L-1} \sum_{m=1}^{M} \bxdg_{lm} \frac{\mathrm{d}^2\phi_{lm}}{\mathrm{d}r^2}(r)$ &  
$\sum_{m=1}^{M} \frac{\bxp_{m+1}+\bxp_{m-1}-2\bxp_{m}}{\dr^2} \chi_{m}(r)$ \\

\hline
\end{tabulary}
\caption{Interpolated solutions ($0$th order derivative) and their $1$st and $2$nd derivatives, for the DG and nodal (NM) representation for the model field at point $r$.  Here $\dr$ is the grid cell width and $\chi_{m}(r)=\max(0,1-\frac{|r_{m}-r|}{\dr})$.  {\label{tab:interpolators}}}
\end{table}

\section{Assimilation in the DG space: can we assimilate denser data?}
{\label{sec:superob}}

The current generation of geophysical DA  systems are restricted in the density of observations that they can effectively assimilate. It has been shown that in standard DA systems, with uncorrelated observational errors, assimilating more observations produces more accurate analyses. However, the improvement becomes marginal when the observational density exceeds $\sim 1$ observation per grid cell \citep{liu_interaction_2002}. On the other hands, spatial correlations between observations are a common occurrence, particularly for satellite data. Such correlations will evidently be present if instrument errors are correlated. But even if they are not, preprocessing and representativiness errors stemming from unresolved processes in the model can introduce spatial correlations between observation errors \citep{janjic_representation_2018,evensen_data_2022}. If these correlations are not accounted for, increasing the observational density is not only ineffective but can even deteriorate DA performance \citep{liu_interaction_2002}. Ideally, any off-diagonal error covariances can be explicitly accommodated in the DA algorithm \citep{miyoshi_estimating_2013, stewart_data_2013, rainwater_benefits_2015, campbell_accounting_2017, fowler_interaction_2018-1,evensen_formulating_2021}. In practice however, efficient minimisation of the cost function in \eq{\ref{eq:cost}} requires diagonal $\bR$ and cross-correlations are eliminated. This is accomplished by discarding data, an operation known as ``thinning'', thus that only  $\sim 1$ observation per cell is assimilated, and/or, by spatially averaging observations, i.e. ``superobbing''. In both cases potentially useful observations are discarded or their information is not properly used. Here we will investigate whether assimilation in DG models with representativeness errors can take advantage from higher-observation densities without having to resort to non-diagonal observation error covariances. 

\subsection{Experimental setup}
{\label{sec:obsSetup}}

The experimental setup is chosen to resemble that in \citet{liu_interaction_2002}: each realisation of the experiment is constructed as follows. 
\begin{enumerate}
    \item In each realisation of the experiment, the ``truth'' $\xtrue$ is assumed to be a realisation from a apriori probability distribution. On its turn, the expectation function $\expectation{\xtrue}$ of this apriori distribution is assumed to be a model field with a spatial mean of zero and spatial autocovariance $cov(\dr)=100 \big(\cos( \frac{8\pi \dr}{\ell})+\frac{4}{3}\sin(\frac{8 \pi \dr}{\ell})\big)e^{-\frac{6 \pi \dr}{\ell}}$ with $\dr$ the distance between points and $\ell=8000 \,\unit{km}$ the width of the periodic domain. I.e. the power spectrum of $\expectation{\xtrue}$ can be written as $\spectrum(\expectation{\xtrue})(\kappa)=|\fourier(cov)|^{2}(\kappa)$ with $\fourier$ the Fourier transform. For each realisation of the experiment, a realisation of the function $\expectation{\xtrue}$ is created from the aforementioned power spectrum following the procedure in \app{\ref{app:generation}} and serves as the mean of the apriori distribution.   
    \item Seventeen ($17$) anomalies are generated by sampling from noise with power spectrum $\spectrum(\kappa) \sim  \kappa^{-\alpha}$, $\spectrum(0)=0$ (zero mean), $\int_{0}^{\infty} \spectrum(\kappa) \mathrm{d}\kappa = 1$ (unit variance), with $\alpha$ depending on the experiment according to the procedure in \app{\ref{app:generation}}. These anomalies are added to the realisation of $\expectation{\xtrue}$ created in the previous step. Keep one of the $17$ members thus created aside as artificial ``truth'' $\xtrue$. The other $N=16$ constitute the background ensemble. I.e. each realisation of the experiments starts off with a different ensemble and different ensemble mean. The ``artificial truth'' accounts for the presence of model error in our experiments.
    \item Sample $\Nobs$ equally-spaced observations from the ``artificial truth'' and add observational noise sampled from a standard normal distribution. The value of $\Nobs$ will vary between experiments. 
    \item Project the $N=16$ members on both the DG space and the nodal space with $M=79$ grid cells. For the nodal space this is done by interpolating the members to the grid cell vertices. The DG coefficients are obtained by integration as outlined in \eq{\ref{eq:legendre_projection}}. 
    \item Sample the members at the observation locations using $\bH^{{\rm NM}}$ for the projection on the nodal space and $\bH^{{\rm DG}}$  for the projection on the DG spaces (see \sect{\ref{sec:da}}). 
    \item Compute the analysis using \eq{\ref{eq:meanUpdate}}.
    \item Calculate the analysis RMSE for the field and some of its derivatives. The RMSE in the $p$th-order derivative is calculated as
    \begin{equation}
        \sqrt{
        \sum_{m=1}^{M} \int_{D_{m}} \big(\mathcal{I}_{p}(r)-\frac{\mathrm{d}^{p} \xtrue}{\mathrm{d}^{p}r}(r)\big)^2 \,\mathrm{d}r
        }
    \end{equation}
    with $\mathcal{I}_{p}$ the interpolator for the $p$th-order derivative in \tab{\ref{tab:interpolators}} corresponding to 
    either the nodal or DG model. 
\end{enumerate}

\subsection{Results observation density}

\begin{figure}[ht!]
\centering
\includegraphics[width=\textwidth,height=.4\textheight, keepaspectratio=true]{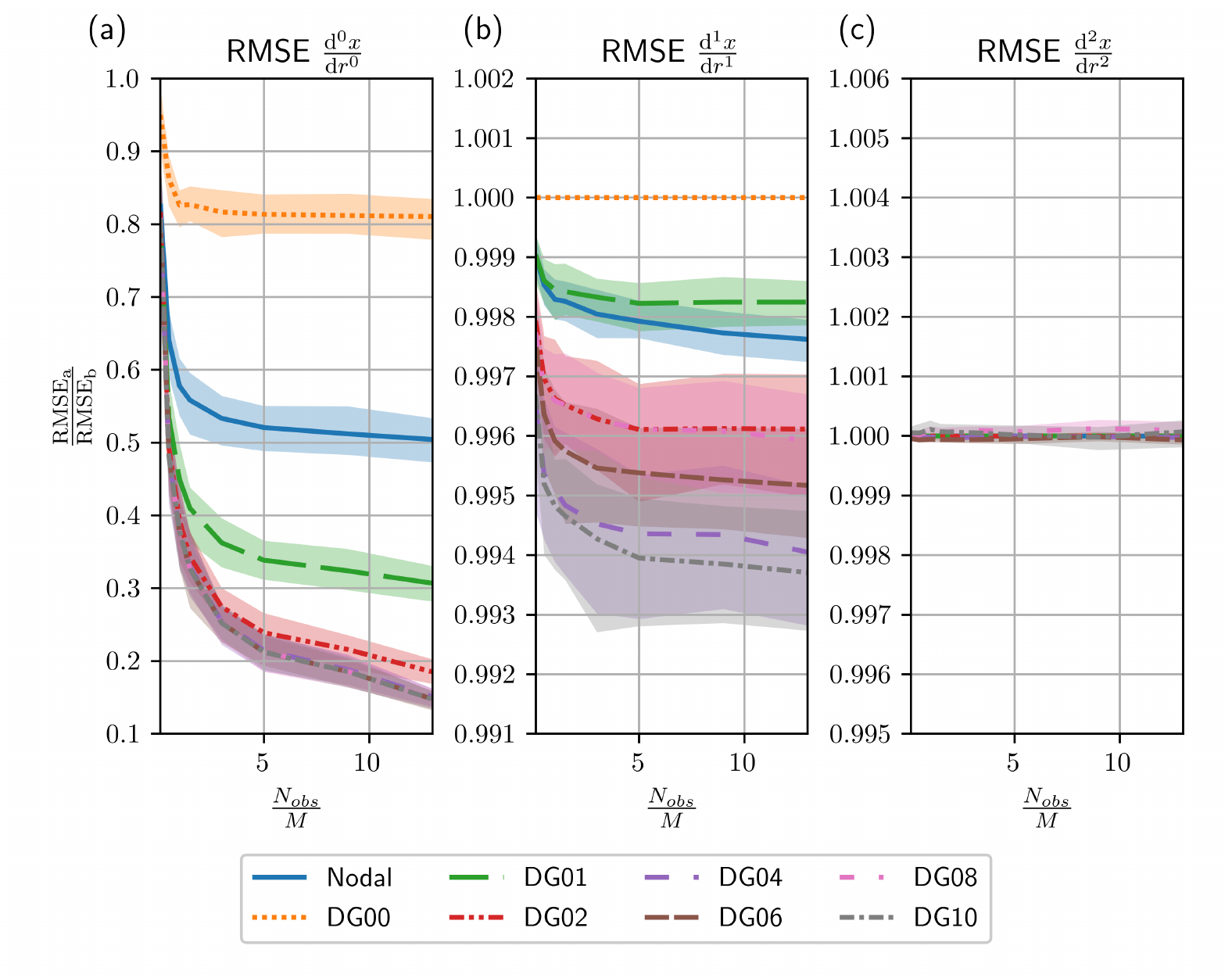}
\caption{Domain-averaged RMS analysis error ($RMSE_{\rm{a}}$)/RMS background error ($RMSE_{\rm{bg}}$) ratio in the (a) the field (b) $1$st-order derivative of the field (c) $2$nd-order derivative of the field as function of the average number of observations per grid cell for the nodal model with $M=79$ grid cells as well as DG models of orders $L-1=0$, $L-1=1$, $L-1=2$, $L-1=4$, $L-1=6$, $L-1=8$ and $L-1=10$, $M=79$ grid cells and background error power spectrum $S$ that scales as $S \sim \kappa^{-4}$. }
{\label{fig:obsDensity1}}
\end{figure}

Following the procedure in \sect{\ref{sec:obsSetup}} using $\Nobs$ observations, we compute domain-averaged forecast (prior to DA) and analysis RMSE for the fields and their first and second order derivatives. Given that both the background and the analysis depend on the space onto which the truth is projected, the ratio between the analysis and background RMSEs is used to quantify the DA improvement. This is then repeated another $49$ times using different ensemble perturbations all created from red noise ($\spectrum({\kappa}) \sim \kappa^{-4}$). After completion, the same 50 ensembles of perturbations are used to carry out the same experiment but now using different numbers of observations $\Nobs$. 

The RMSEs from these $50$ experiments and their $90\%$-confidence intervals calculated using percentile bootstrap \citep{efron_introduction_1994} are shown in \fig{\ref{fig:obsDensity1}}a as function of the average number of observations per grid cell. For the nodal model the decrease of the ratio with increasing observation densities starts to level off at $\frac{\Nobs}{M} \approx 1.5$ with $M=79$ the number of grid cells. For higher-order DG models, i.e. DG models using polynomials of order $\geq 1$, the RMSE ratio decreases faster with increasing observation density than for the nodal case. The difference in slope is most pronounced when  $\frac{\Nobs}{M} \leq 3$, but the RMSE in the DG models continues to decrease even after this point. As a result, the DA in the higher-order DG models outperforms the DA in the nodal model significantly at the $90\%$-level. This behaviour, i.e. the capability to assimilate and benefit from denser data, is a consequence of the DG models having more degrees of freedom than the nodal model and can therefore provide a better fit to the observations. More surprisingly is that for the $0$th-order derivative (i.e. the model field itself) there is no DA benefit beyond a DG order of $4$. This can be explained by the fact that the background error spectrum in the experiment is dominated by $\spectrum({\kappa}) \sim \kappa^{-4}$. I.e. it is dominated by the large scales (small $\kappa$). Higher-order polynomials represent smaller scales and consequently the background variance explained by the higher-order DG coefficients is so small that effectively no DA updates are made to these coefficients. Contrary to \citet{liu_interaction_2002} no increase of RMSE with increasing observational density was detected. This because the observational error correlations introduced by the representativeness errors in this experiment are small compared to observational error correlations introduced in\citet{liu_interaction_2002}. 

The RMSE ratios for the $1$st derivative decrease (i.e. improve) with increasing DG order as shown in \fig{\ref{fig:obsDensity1}}b. The improvement over the nodal model is significant at the $90\%$-confidence level for higher-order DG models. Furthermore, the relative improvement increases consistently with the DG order. However, the actual size of this improvement is rather marginal ($<0.7\%$; cf. the y-axis ranges in \fig{\ref{fig:obsDensity1}}a and \fig{\ref{fig:obsDensity1}}b). Finally, \fig{\ref{fig:obsDensity1}}c shows that DA has no significant impact on the errors in the $2$nd derivative. 

\begin{figure}[!ht]
\centering
\includegraphics[width=\textwidth,height=.4\textheight,keepaspectratio=true]{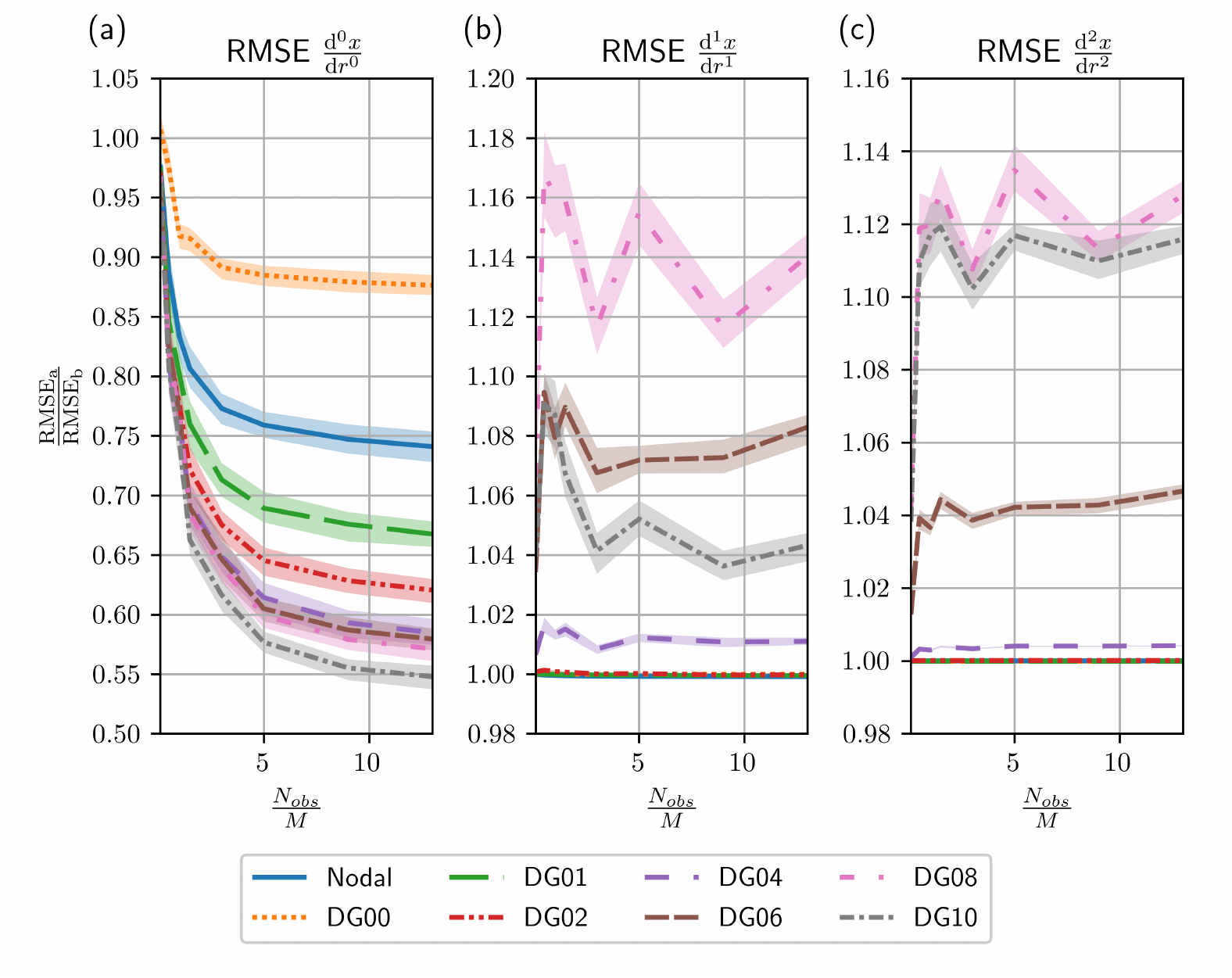}
\caption{As \fig{\ref{fig:obsDensity1}} but now with a background error power spectrum $S$ that scales as $S \sim \kappa^{-1}$. }
{\label{fig:obsDensity2}}
\end{figure}

The absence of improvement when the DG order is increased beyond $4$ suggests that the potential of DG-based DA to improve over nodal (i.e. standard) DA is related to (effectively bounded by) the scales actually present in the background error. To investigate this behaviour, we perform the same type of experiments but this time with background errors sampled from a pink noise spectrum, i.e. spectral slope $\alpha=-1$, that by construction includes smaller spatial scales. Results are shown in \fig{\ref{fig:obsDensity2}}. In particular, 
\fig{\ref{fig:obsDensity2}}a confirms the finding in \fig{\ref{fig:obsDensity1}} that the higher-order DG models outperform the nodal model. Nevertheless, in contrast to \fig{\ref{fig:obsDensity1}}a there is now a consistent reduction of the RMSE ratios with increasing DG order. It is worth noting however that the RMSE ratios are higher than those obtained with a spectral slope of $\alpha=-4$ (cf. \fig{\ref{fig:obsDensity1}}a). The reason for this will be discussed in more detail at the end of this section. For the $1$st and $2$nd derivatives (\fig{\ref{fig:obsDensity2}a} and \ref{fig:obsDensity2}b) the situation is substantially different from that in \fig{\ref{fig:obsDensity1}} with the RMSE ratios increasing as the DG order increases: DA deteriorates the representation of the gradient and the Laplacian of the solution. 

\begin{figure}[ht!]
\centering
\includegraphics[width=\textwidth,height=0.6\textheight, keepaspectratio=true]{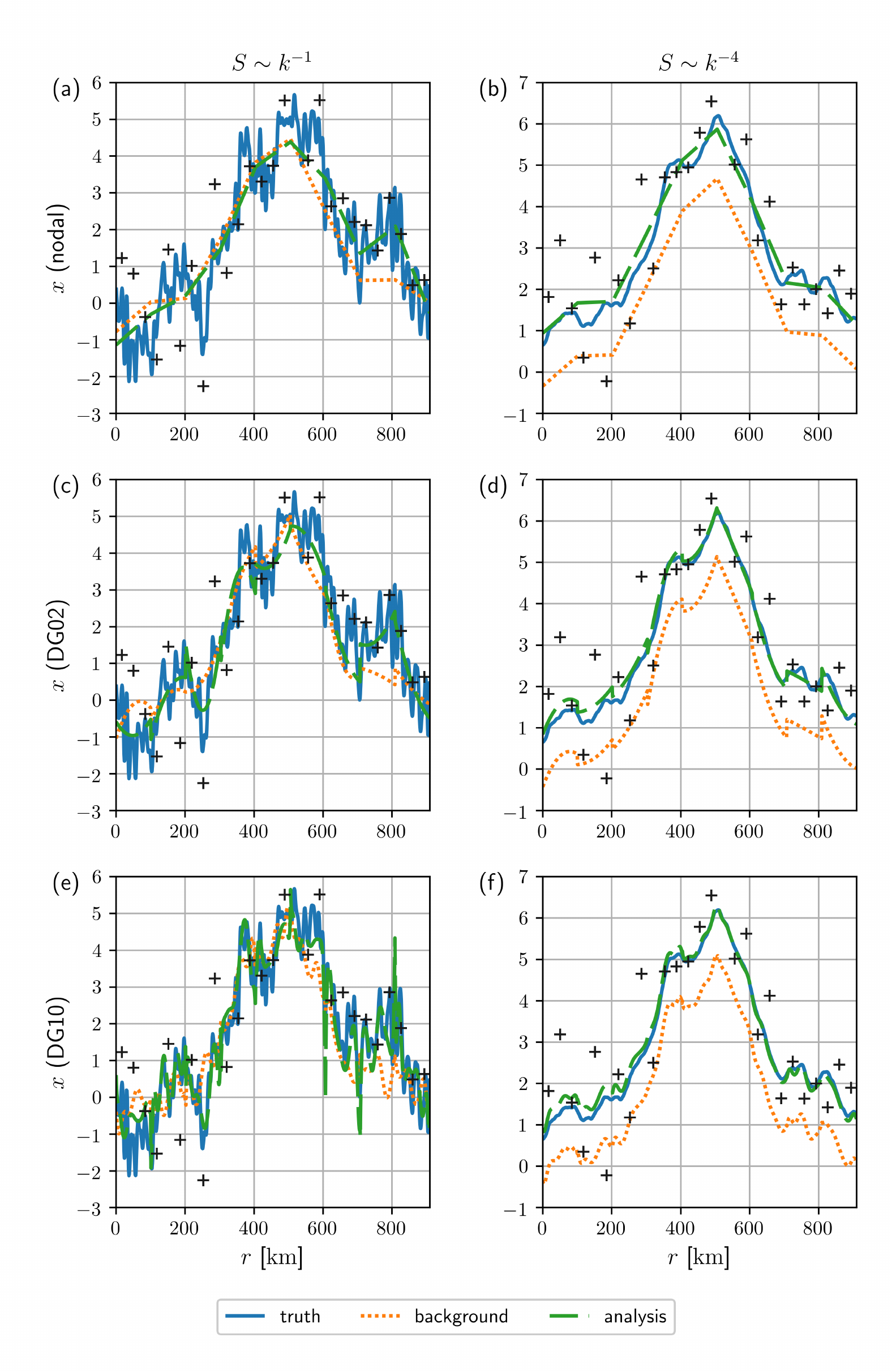}
\caption{Example of the the (blue) the true field, (orange) the background estimate, and (green) analysis for (top row) the nodal model, (centre row) the DG order-2 model, (bottom row) the DG order-10 model with background error sampled from a spectrum with (left column) $\alpha=-1$ and (right column) $\alpha=-4$. On average 3 observations are assimilated per grid cell. The observed values are depicted as black $+$-signs. Only a limited part of the domain (9 grid cells) is shown. }
{\label{fig:obsFields}}
\end{figure}

To understand and illustrate why this occurs, we plot in \fig{\ref{fig:obsFields}} the observations, the truth, the background and the analysis for DG orders $2$ and $10$, the nodal solution, and the background error for the two noise spectra under consideration. In the case of a pink spectrum ($\alpha=-1$, panels (a), (c) and (e)), the spectrum of the background error in the $1$st ($2$nd) derivative has a slope of $\alpha=1$ ($\alpha=3$), which follows a blue spectrum. This implies that most of the error in the $1$st and $2$nd derivatives is in the smaller scales. As the dashed orange line (the background estimate) in \fig{\ref{fig:obsFields}}a,c shows, this small-scale portion of the spectrum is filtered out by projecting on the nodal or on the DG02 space. Therefore there is not much error left to be removed by the DA and the RMSEs ratios are of $\sim 1$ (cf. \fig{\ref{fig:obsDensity2}}b,c). The higher-order polynomials in the DG10 model contain smaller scales. The freedom contained in these scales is used to fit the observations (see \fig{\ref{fig:obsFields}}e). However, to fit the observations the analysis field must make sharp turns (see e.g. \fig{\ref{fig:obsFields}}e near position $100 \,\unit{km}$ and $700 \,\unit{km}$) leading to large oscillations in the analysis, especially near the cell boundaries. This behaviour is similar the Runge phenomenon  observed for the interpolation polynomial: as the DG order increases the analysis starts to fit the observations. However, increases in order do not convergence uniformly to some limit. Instead near the boundaries oscillations develop of which the amplitude increases with increasing order. These oscillations can result in point-wise divergence (see e.g.  \fig{\ref{fig:obsFields}}e near $700 \,\unit{km}$). 
As the scale of these oscillations is small, their contribution to the RMSE is more noticeable for the derivatives than for the field itself.  In summary, for high-order DG models the reduction of errors in the field introduces errors in the small scales that are magnified by differentiation. 

With a red error spectrum ($\alpha=-4$) the spatial correlations in the background errors are longer. For portions of the domain smaller than the correlation scale this gives the impression that the background is the truth plus a fixed offset (see \fig{\ref{fig:obsFields}}b,d,f) and errors in its derivatives are small. Since most of the error is located in the large scales, background estimates for the derivative are already good and little correction from DA can be achieved. 

The experiment has been repeated (not shown) using observational error standard deviations of $\sigma_{o}=0.5$ ($\sigma_{o}=1.5$). In these cases the analysis RMS error ratio is lower (higher) than those in \fig{\ref{fig:obsDensity1}} and \fig{\ref{fig:obsDensity2}}, but qualitatively there is no difference. 

\begin{figure}[ht!]
\centering
\includegraphics[width=\textwidth,height=0.6\textheight, keepaspectratio=true]{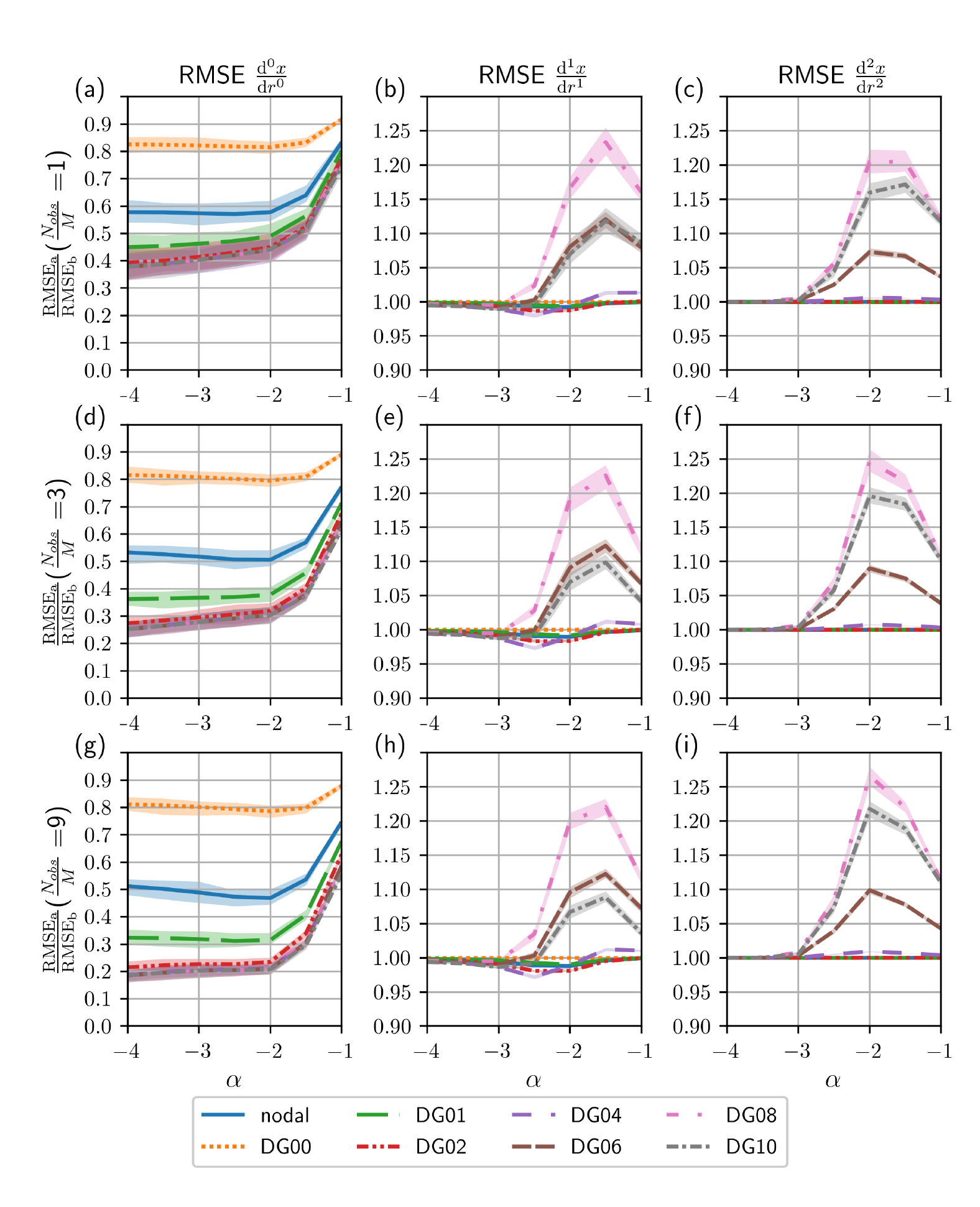}
\caption{Investigation into the dependence of the RMSE on the scales in the background error. Shown are the analysis/background-RMSE ratios of the (left column) $0$th, (centre column) $1$st and (right column) $2$nd derivatives as function of the slope $\alpha$ of the power spectrum for experiments assimilating on average (top row) $\frac{\Nobs}{M}=1$, (centre row) $\frac{\Nobs}{M}=3$ and (bottom row) $\frac{\Nobs}{M}=9$ observations per grid cell.}
{\label{fig:obsSlope}}
\end{figure}

The dependence of the DA performance on the background error spectrum and its spatial scales is further explored in \fig{\ref{fig:obsSlope}}. In \fig{\ref{fig:obsSlope}} the RMSE ratio for the model field and its derivatives is shown as function of the slope of the power spectrum $\alpha$ used to generate the background errors three different observational densities. The left column of \fig{\ref{fig:obsSlope}} shows that for the model field itself DA performance improves in all DG orders as the spectrum from which the background errors are sampled becomes redder (lower $\alpha$): the smaller the dominant scales in the error are, the smaller the improvement obtained using DG-based DA. However, models in which the smaller scales dominate the background errors benefit the most from increasing the DG order.  The centre and right column of \fig{\ref{fig:obsSlope}} show that the negative impact of DA on the derivatives in the higher-order DG models starts to disappear when the background error spectrum slope drops below $\alpha < -2$. 

\section{DG-based scale-dependent localisation}
{\label{sec:localisation}}

In this section we shall consider exclusively DG models using Legendre polynomials as basis function. We will look at their filtering properties and will show how these properties can be used to introduce a scale-dependent localisation for ensemble-based DA methods. 

\subsection{Filtering Properties of Legendre Polynomials}
{\label{sec:filtering}}

The Legendre polynomials form a family of polynomials of increasing order, $l$, on domain $\Dref=[-1,1]$ given by the recursion relation \citep{beals_special_2016}
\begin{eqnarray}
    \legendre{l}(r) = 
    \begin{cases}
        1 \quad \text{ if } l=0\\
        r \quad \text{ if } l=1\\ 
        \frac{2l-1}{l}r \legendre{l-1}(r)-\frac{l-1}{l}\legendre{l-2}(r) \quad \text{ if } l>1 .
    \end{cases}
\end{eqnarray}

\begin{figure}[h!]
    \centering
    \includegraphics[width=\textwidth, height=.3\textheight, keepaspectratio=true]{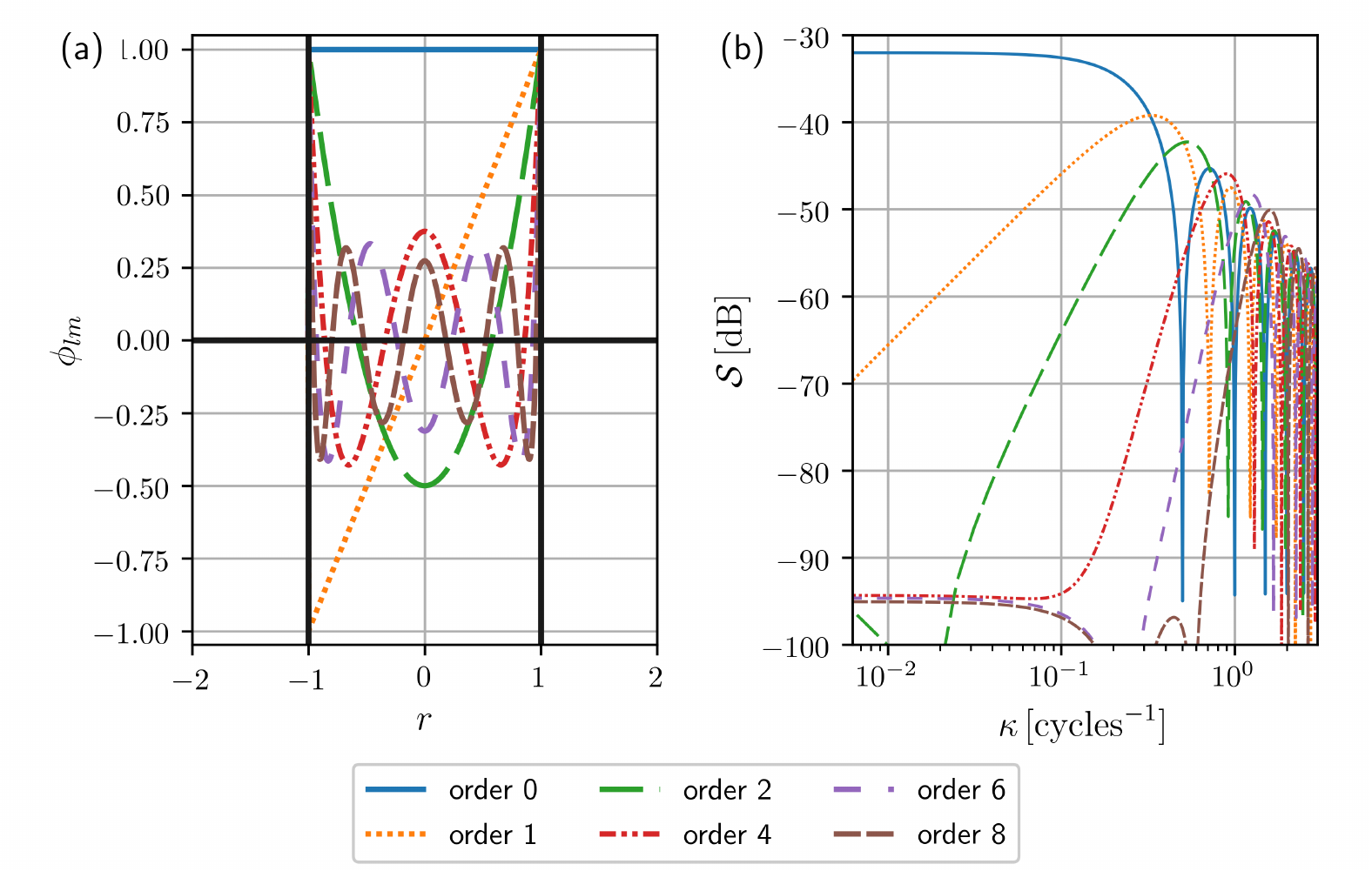}
    \caption{(a) Examples of Legendre basis polynomials $\phi_{lm}$ of different polynomial orders $l$ on a domain consisting of $79$ cells of size $2$. Shown examples all have their support in the grid cell $[-1,1]$. (b) Power spectrum of the polynomials shown in (a). {\label{fig:legendre_polynomials}}} 
\end{figure}

The Legendre polynomials are orthogonal, i.e. $\int_{\Dref} \legendre{l}(r) \legendre{l'}(r) \,dr = \frac{2}{2l+1} \delta_{ll'}$. Examples of the basis functions $\phi_{klm}$ based on the Legendre polynomials are shown in \fig{\ref{fig:legendre_polynomials}a}.

The power spectra in \fig{\ref{fig:legendre_polynomials}b} show that as the polynomial order increases, a relatively larger part of the spectral power is contained in the smaller scales ($\kappa > 1$) while less-and-less ends up in the larger scales ($\kappa \sim 0$). This suggests that the Legendre polynomials can be used to construct a band-pass filter. In the following we will describe how such a filter can be constructed. For simplicity, we will assume that the model grid $D$ consists of grid cells of width $2$. Such a situation can always be created by applying a suitable coordinate transformation to the model's differential equations before integrating the model. After such a transformation the transform function $\Psi_{m}:\Dref \to D_m$ simplifies to $\Psi_{m}(r)=r_{m}-r$ where $r_{m}$ is the centre of the $m$th grid cell. This transformation can immediately be generalized to a continuum as $\Psi_{r}(r')=r-r'$. Furthermore, we extend the definition of the Legendre polynomials $\refBase_{l}$ from $\Dref$ to $\R$ as
\begin{equation}
    \underline{\refBase}_{l}(r') = 
    \begin{cases}
    \refBase_{l}(r') & \text{ if } r' \in \Dref \\
    0 & \text{otherwise} ,
    \end{cases}
\end{equation}
and use this to define a filter $\bandpass_{l}$
\begin{equation}
    \bandpass_{l}(x)(r,t) 
    \overset{def}{=} (x * \underline{\refBase}_{l})(r) 
    \overset{def}{=}\int_{\R} \underline{\refBase}_{l}(r-r') x(r',t) \,{\mathrm d}r' 
    =\int_{-1}^{1} \refBase_{l}(r') x(\Psi_{r}(r'),t) \,{\mathrm d}r' ,
    \label{eq:convolution}
\end{equation}
where $*$ denotes convolution. From the convolution theorem it now follows that
\begin{equation}
     |\fourier\bandpass_{l}(x)|^2(\kappa) \overset{def}{=}|\fourier(x * \refBase_{l})|^2(\kappa)=|\fourier(x \circ \Psi_{r})|^2(\kappa)
    |\fourier(\refBase_{l})|^2(\kappa)=|\fourier(x)|^2(\kappa)
    |\fourier(\refBase_{l})|^2(\kappa) ,
    \label{ref:spectrum}
\end{equation}
where $\fourier$ is the Fourier transform, $\kappa$ is the wavenumber and by definition the power spectrum reads $\spectrum(\refBase_{l})=|\fourier(\refBase_{l})|^2$. I.e. the filter $\bandpass_{l}$ modifies the spectral power contained in $x$ at wavenumber $\kappa$ with a factor $|\fourier(\refBase_{l})|^2(\kappa)$. The power spectra  $|\fourier(\refBase_{l})|^2(\kappa)$ in \fig{\ref{fig:legendre_polynomials}b} show that the peak of this spectrum shifts to higher wavenumbers as $l$ increases. Consequently, $\bandpass_{l}$ removes spectral power at low wavenumbers (long scales) and the range of wavenumbers (scales) over which this removal takes place increases with increasing $l$.  Furthermore, using the orthogonality of the Legendre polynomials and $\Psi_{r_{m}}=\Psi_{m}$ we find that $\bandpass_{l}(x)(r_{m},t)=\frac{2}{2l+1}x_{lm}(t)$. Therefore, apart from a scaling constant, the DG coefficients associated with the $l$th-order Legendre polynomial are precisely the values of the band-pass filtered field at the grid cell centres.

\subsection{Localisation Matrix}
{\label{sec:locMatrix}}

In practical applications, the size of the ensemble used to estimate $\bB$ in \eq{\ref{eq:meanUpdate}} and \eq{\ref{eq:ensUpdate}} is orders of magnitude smaller than the dimension of the model space. This introduces a sampling error in the estimate $\bB$. For correlations close to zero this sample error can overshadow the true correlation. These spurious correlations are suppressed using localisation. As discussed in \sect{\ref{sec:intro}}, covariance localisation is to be preferred over domain localisation when dealing with dense (satellite) observations. Covariance localisation replaces $\bB$ with $\bB \circ \bL$, the Schur-, or element-wise, product of $\bB$ with a positive-definite, symmetric matrix $\bL$. Two types of methods are in common use to construct $\bL$: 1) parameterised schemes in which the elements of $\bL$ are assumed to depend on a very limited number of parameters. Parameter values are obtained from calibration experiments by minimising  metrics like RMSE. The best known scheme within this category is GC-scheme \citep{gaspari_construction_1999-2} in which the elements of $\bL$ depend on a single parameter, the localisation length scale. Other examples belonging to this category can be found in \citet{anderson_exploring_2007, moosavi_machine_2018}. 2) non-parameterised schemes in which localisation factors are obtained as the expectation values of unknown distributions \citep{anderson_exploring_2007, anderson_localization_2012, flowerdew_towards_2015, menetrier_linear_2015}. These expectation values are approximated from the model under ergodicity assumptions. 

Regardless of the method used, the optimal parameter values/the expectation values will depend on the signal-to-noise ratio in the sample covariance. In particular, it may be expected that if the length scales in the background errors change so does the distance at which noise starts to exceed the signal: the length scale in $\bL$. Traditionally, localisation factors $\bL$ have been estimated assuming a single length scale without considering the multi-scale feature of the model fields. If multiple scales are present in the field this can result in localisation factors that suppress true correlations at one scale, but fail to suppress the noise at other scales. In \sect{\ref{sec:filtering}} we showed that the Legendre polynomials act as a band-pass filter separating the the model field by length scale. This allows to circumvent aforementioned problem by applying different amounts of localisation to different scales in the signal. More formally, define 
\begin{equation}
    \bL^{{\rm DG}} = \sum_{l=1}^{L-1}\sum_{l'=1}^{L-1} (\projection^{{\rm DG}}_{l})\T \bL^{{\rm DG}}_{ll'} \projection^{{\rm DG}}_{l'} , 
    \label{eq:scale_L}
\end{equation}
where
\begin{equation}
\projection^{\rm{DG}}_{l} = \sum_{m=1}^{M} \contravec_{m} \otimes ( \bigotimes_{l'=0}^{L-1}  \delta_{ll'} \covec_{m})
\end{equation}
is the projection operator that selects only those elements from $\bx$ that are associated with the $l$th basis function. Here $\bL_{ll'}$ only contains the localisation factors for the covariance between DG coefficients associated with Legendre polynomial $l$ and $l'$. 

As $\bL^{^{\rm DG}}$ in \eq{\ref{eq:scale_L}} requires the construction of multiple localisation matrices $\bL^{{\rm DG}}_{ll'}$ the amount of calibration required makes application of the parametrised localisation scheme impractical. Instead we construct it as $\bL^{{\rm DG}}_{lml'm'}=\bL^{{\rm DG}}_{ll'}(\sin(\pi \frac{|r_{m}-r_{m'}|}{\ell}))$ where the choice of the sinus stems from the use of a periodic domain. The ``kernel'' matrix $\bL^{{\rm DG}}_{ll'}$ is obtained using the optimal Schur filter \citep{menetrier_linear_2015}. 

For the sake of completeness we will summarise the optimal Schur filter here. We start from \eq{64} in \citet{menetrier_linear_2015}:
\begin{equation}
    \bL^{\rm{DG}}_{lml'm'} = \frac{N-1}{(N-2)(N+1)}(N-1 - \frac{
    \expectation{\sum_{n=1}^N \ba_{lm}^{(n)}\ba_{lm}^{(n)}
    \sum_{n'=1}^{N} \ba_{l'm'}^{(n')} \ba_{l'm'}^{(n')} }
    }{
    \expectation{\sum_{n=1}^{N}  \ba_{lm}^{(n)} \ba_{l'm'}^{(n)} \sum_{n'=1}^{N} \ba_{lm}^{(n')} \ba_{l'm'}^{(n')}}
    }
    )
    \label{eq:optimal_L}
\end{equation}
Here $\expectation{\cdot}$ is the expectation value of $\cdot$ and $\ba$ the ensemble perturbation. An ergodicity assumption is made and the expectation values are approximated using their spatial averages  
\begin{eqnarray}
    && \expectation{\sum_{n=1}^{N} \ba_{lm}^{(n)}\ba_{lm}^{(n)}
    \sum_{n'=1}^{N} \ba_{l'm'}^{(n')} \ba_{l'm'}^{(n')}} \approx
      \mathbf{v}_{ll'}\left[ \frac{\pi |r_{m}-r_{m'}|}{\ell}) \right] \nonumber \\ && \overset{\rm{def}}{=} 
    \frac{1}{2M}\sum_{i=1}^{M} 
    ( \sum_{n=1}^{N} \ba_{li}^{(n)} \ba_{li}^{(n)} )
    \left[ \sum_{n'=1}^{N} (\trans_{m-m'}\ba_{l'}^{(n')})_{i} (\trans_{m-m'}\ba_{l'}^{(n')})_{i} +
    \sum_{n'=1}^{N} (\trans_{m'-m}\ba_{l'}^{(n')})_{i} (\trans_{m'-m}\ba_{l'}^{(n')})_{i} \right]  \nonumber \\ && =
    \frac{1}{2}\Big[ \big(\sum_{n=1}^{N} \ba_{l}^{(n)} \circ \ba_{l}^{(n)} \big) 
    * \big(\sum_{n'=1}^{N} \mirror\ba_{l'}^{(n')} \circ \mirror\ba_{l'}^{(n')} \big) \Big]_{m-m'} \nonumber \\ && + 
    \frac{1}{2}\Big[ \big(\sum_{n=1}^{N} \ba_{l}^{(n)} \circ \ba_{l}^{(n)} \big) 
    * \big(\sum_{n'=1}^{N} \mirror\ba_{l}^{(n')} \circ \mirror\ba_{l'}^{(n')} \big) \Big]_{m'-m}
    \label{eq:vv}
\end{eqnarray}
and 
\begin{eqnarray}
&& \expectation{\sum_{n=1}^{N} \sum_{n'=1}^{N} \ba_{lm}^{(n)} \ba_{l'm'}^{(n)} \ba_{lm}^{(n')} \ba_{k'l'm'}^{(n')}} \approx
 c_{ll'} \left[ \sin( \frac{\pi |r_{m}-r_{m'}|}{\ell}) \right] \nonumber \\ &&
\overset{\rm{def}}{=} \frac{1}{2M} \sum_{n'=1}^{N} \sum_{n=1}^{N} \sum_{i=1}^{M} \ba_{li}^{(n)} (\trans_{m-m'}\ba_{l'}^{(n)}+\trans_{m'-m}\ba_{l'}^{(n)})_{i} \ba_{li}^{(n')} (\trans_{m-m'}\ba_{l'}^{(n')}+\trans_{m'-m}\ba_{l'}^{(n')})_{i} \nonumber \\
&&= \frac{1}{2}\sum_{n=1}^{N} \sum_{n'=1}^{N} \big( \ba^{(n)}_{l} * \mirror \ba^{(n)}_{l'} \big)_{m-m'} \big( \ba^{(n')}_{l} * \mirror \ba^{(n')}_{l'} \big)_{m-m'} \nonumber \\
&&+ \frac{1}{2}\sum_{n=1}^{N} \sum_{n'=1}^{N} \big( \ba^{(n)}_{l} * \mirror \ba^{(n)}_{l'} \big)_{m'-m} \big( \ba^{(n')}_{l} * \mirror \ba^{(n')}_{l'} \big)_{m'-m}
\label{eq:B2}
\end{eqnarray}
where $\trans_{j}$ is a shift of the state by $j$ grid points on a periodic grid, $\mirror$ the mirror operator on a periodic domain and $*$ given by
$
(\mathbf{a} * \mathbf{b})_{i} = \frac{1}{M}\sum_{m=1}^{M} \mathbf{a}_{m} (\trans_{i} \mirror \mathbf{b})_{m}
$
the discrete convolution operator for a periodic domain. The resulting expression for the localisation coefficients is
\begin{equation}
    \bL^{\rm{DG}}_{lml'm'} \approx \mathbf{L}^{DG}_{ll'}\big(\sin(\frac{\pi|r_{m}-r_{m'}|}{\ell})\big) = 
    \frac{N-1}{(N-2)(N+1)}\Big(N-1 - \frac{v_{ll'}\big(\sin(\frac{\pi|r_{m}-r_{m'}|}{\ell})\big)}{c_{ll'}\big(\sin(\frac{\pi|r_{m}-r_{m'}|}{\ell})\big)}
    \Big).
\end{equation}

The expression for the nodal case can be found by setting $L=1$ and 
omitting any indices $l$, $l'$. A comparison of the computational cost of the scale-dependent method versus the localisation operator for the nodal case can be found in appendix~\ref{app:locCost}. 

\subsection{Setup Localisation Experiments}
{\label{sec:locComparison}}

In order to compare the differences between $\bB$ created using scale-dependent localisation, standard scale-independent spatial localisation and no localisation, we generate a large ensemble of $10,000$ univariate ensemble members as Fourier series following the procedure outlined in appendix~\ref{app:generation} using $J=395$ degrees of freedom in the Fourier coefficients, a pink noise spectrum $\spectrum(k) \sim \kappa^{\alpha}$, $\alpha=-1$ and with a standard deviation of 1. Covariance estimated from this $10,000$-member ensemble is assumed sufficiently large to have negligible sampling error. 

\begin{figure}[ht!]
    \centering
    \includegraphics[width=\textwidth]{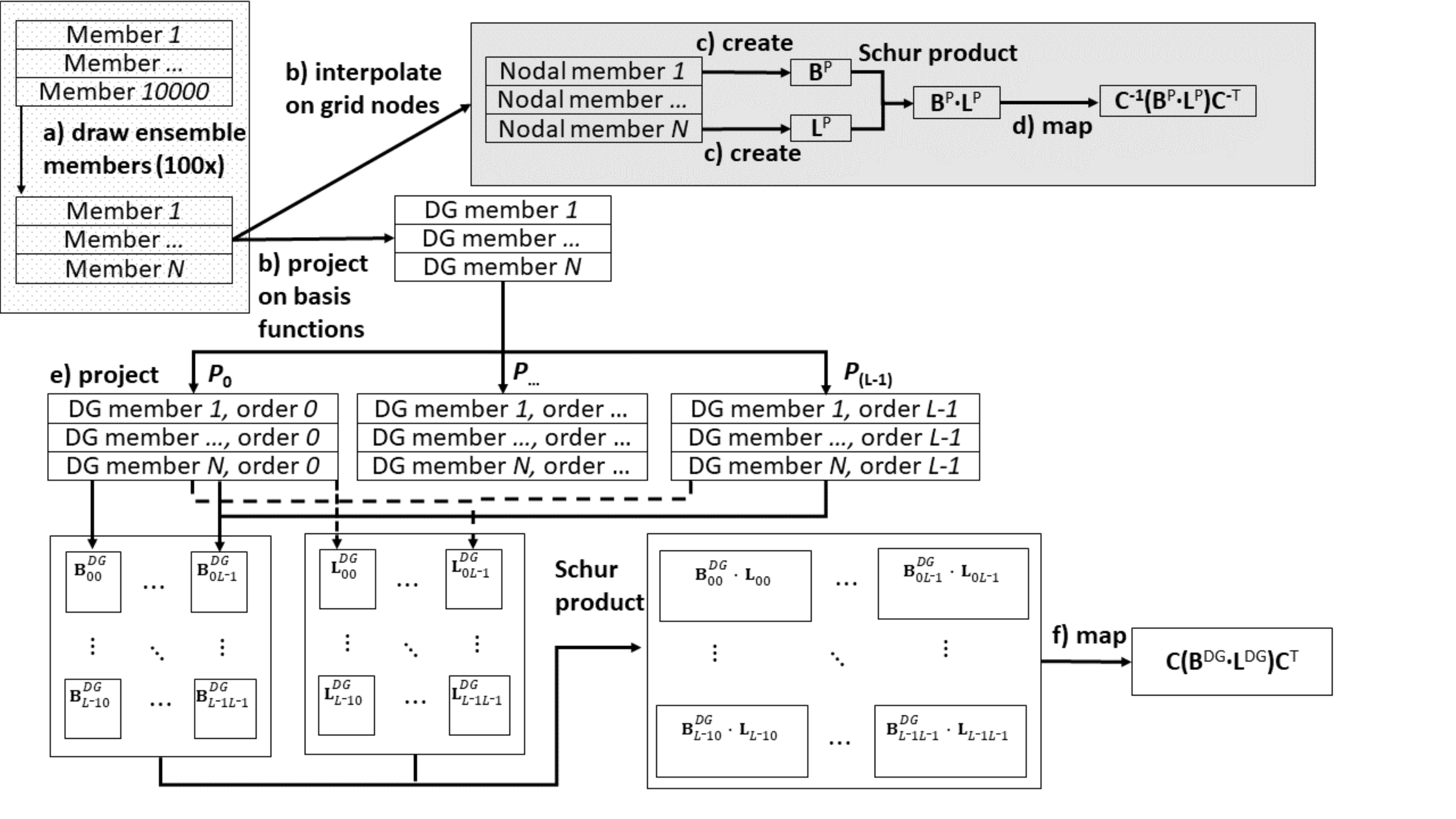}
    \caption{Schematic overview of the experimental setup to create a localised covariance in nodal space (dotted grey and solid grey background) and in DG space (dotted grey and white background). A large $10,000$-member ensemble is created, (a) from this ensemble members are randomly selected to create smaller subensembles, (b) the subensemble members are interpolated/projected to create {\it nodal/DG states}, (c) from the \textit{nodal states} covariance and localisation operators are created and (d) the localised covariance operator is mapped to DG space. Simultaneously, (e) the \textit{DG states} are split by the order of their coefficients and for each pair of orders covariances and localisation operators are calculated.(f) The resulting localised covariances are mapped from DG to nodal space or vice-versa using the linear transform $\coordTrans$ defined in \eq{\ref{eq:coordTrans}}. Such a transformation is only possible in this section as the number of nodal points is equal to the dimension of the DG space. 
    {\label{fig:scalelocalization}}}
\end{figure}

The ensemble members generated are processed in two different ways.
\begin{enumerate}
\item Continuing the procedure outlined in appendix~\ref{app:generation}, each ensemble member is projected onto a 4th-order Legendre basis ($L-1=4$) defined on a periodic domain with length $\ell=8000 \,\unit{km}$ and $M^{\rm{DG}}=79$ equispaced grid cells. 
\item In the nodal approach each ensemble member is interpolated onto an equispaced grid. This  grid will contain $M^{\rm{NM}}=M^{\rm{DG}} L=J$ grid points so that a state on this grid has the same dimensionality as the \textit{DG state} and the field in Fourier space. The \textit{nodal state} on this grid is obtained by interpolating the ensemble member expressed as Fourier series onto the grid nodes. 
\end{enumerate}
Subensembles of size $N$ with sampling error are created by randomly selecting $N$ ensemble members from $10,000$-member ensemble and storing their representation as \textit{DG state} and \textit{nodal state}.
A schematic overview of the whole process is given in \fig{\ref{fig:scalelocalization}}. 

\subsection{Resulting Localisation Factors}
{\label{sec:locCoefficients}}

\begin{figure}[ht!]
\centering 
\includegraphics[width=\textwidth,height=.35\textheight,keepaspectratio=true]{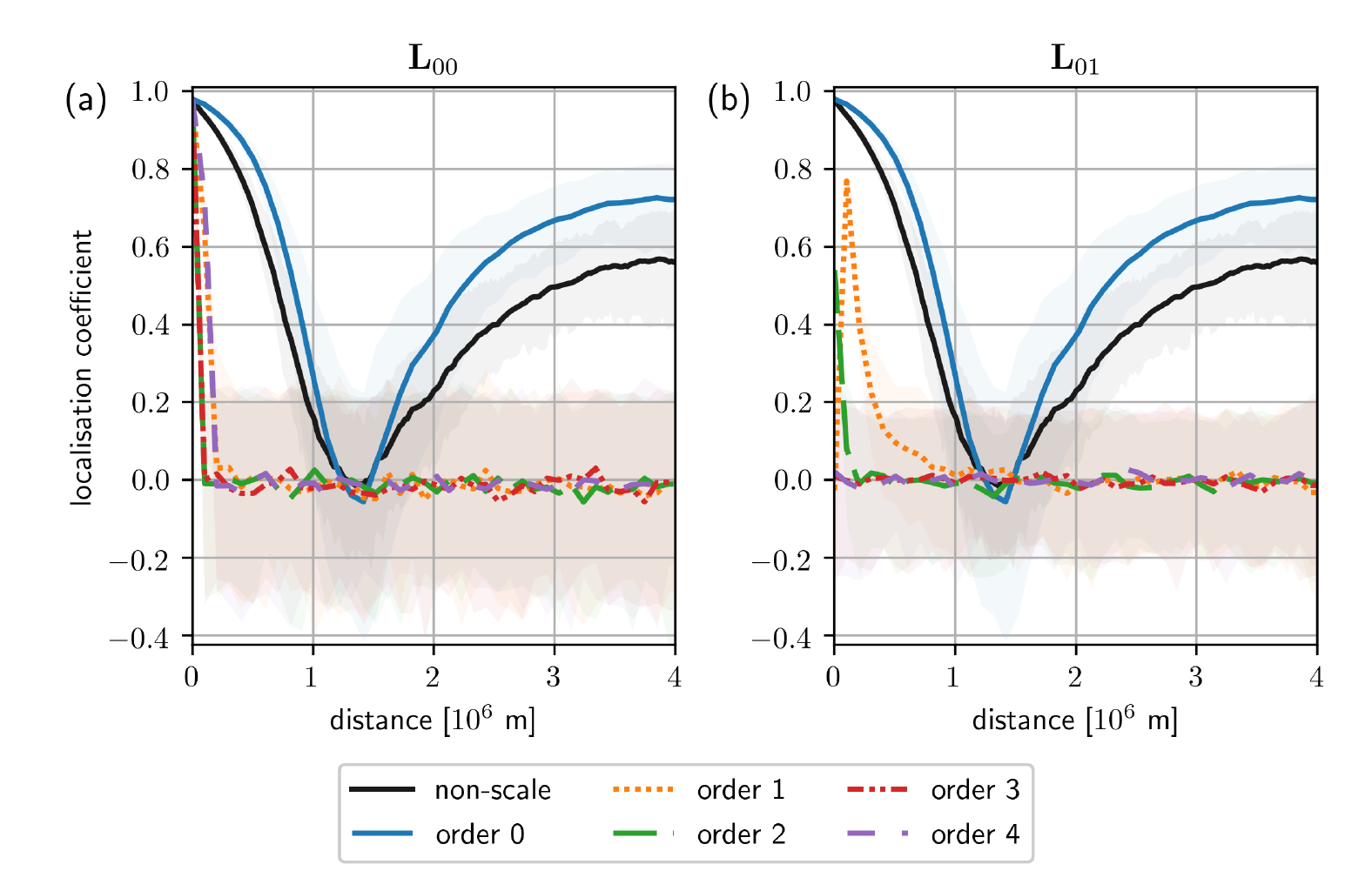}
\caption{(a) Localisation coefficients (colours) $\bL^{\rm DG}_{ll}$ for different orders $l$ as well as (black) $\bL^{\rm NM}$ as function of distance between grid cells based on $N=96$-member ensemble. (b) As a) but the colours are now showing cross-order coefficients $\bL^{\rm DG}_{0l}$ between the $0$th-order coefficient and coefficients of higher orders; the black line is the same in (a) and (b). {\label{fig:locCoef96}}}
\end{figure}

First we verify that localisation distances are indeed scale-dependent. To this end, $N=96$ \textit{DG states} are randomly selected from the $10,000$-member ensemble. The localisation tensor $\bL^{{\rm DG}}$ is calculated from these members using \eq{\ref{eq:optimal_L}}. As a result of the ergodicity assumption in \eq{\ref{eq:vv}} and \eq{\ref{eq:B2}} the localisation factors $\bL_{lml'm'}$ are solely a function of the orders, $l$ and $l'$, of the associated polynomials and of the difference in grid cell indices $|m-m'|$. For equispaced grids, as the one used in our experiments, the latter can be related to distance between points via $\frac{\ell}{M}|m-m'|$. 
For the non-scale dependent localisation factors, $\bL^{{\rm NM}}$, we use again \eq{\ref{eq:optimal_L}} on the \textit{nodal states} created from the same $96$ ensemble members. The calculation of both $\bL^{{\rm DG}}$ and $\bL^{{\rm NM}}$ is repeated $100$ times using different $96$-member ensembles and for each localisation coefficient the mean, $5\%$- and $95\%$-percentile over these $100$ repetitions are shown in  \fig{\ref{fig:locCoef96}}. 

The black line in \fig{\ref{fig:locCoef96}}a,b shows the localisation factor for the {\it nodal state} as function of distance between grid position. As expected the maximum localisation factors can be found at zero distance. Nevertheless, the inspection of the non-scale case (black curve) reveals a few aspects deserving clarification. First, contrary to many other localisation schemes, the maximum localisation  factor is smaller than $1$ because the optimal Schur localisation scheme not only corrects for the sampling error in correlation but also for that in the sample variance \citep{menetrier_linear_2015}. 
As the spectrum of the ensemble perturbations scales as $\kappa^{-\alpha}$ the 1st Fourier mode dominates the perturbations. As this mode is a sinusoid with wavelength $\ell$, the autocorrelation as function of distance between points first decrease towards zero and then becomes negative, whilst its absolute value increases, as distance approaches $\frac{1}{2}\ell$. As the localisation coefficient is a function of the absolute value of the autocorrelation, it also initially decreases towards zero at $1.5\cdot10^{6} \,\unit{m}$ but it starts to increase again as the distance approaches $\frac{1}{2}\ell$. 

The localisation factor $\bL^{\rm{DG}}_{0m0m'}$ (blue) closely resembles the non-scale dependent localisation but with higher values thus suppressing the ensemble covariance less than its non-scale dependent counterpart. The behaviour for the higher-order DG localisation factors is significantly different as is visible in \fig{\ref{fig:locCoef96}}a. Within 1 grid cell length the localisation factor drops from its maximum value of 0.97 to its lower bounds which varies around 0.0. 
Cross-order DG localisation factors in \fig{\ref{fig:locCoef96}}b also show rapid decrease to values around $0.0$, with cross-order localisation factors between $0$th-order coefficients and coefficients of order 3 (red) and higher (purple) exhibiting no significant deviation from zero. Overall, \fig{\ref{fig:locCoef96}} shows that localisation length scale, i.e. the length scale over which $\bL_{lml'm'}$ goes to zero, strongly depends on the order of the coefficients $l$, $l'$ being localised. In particular, the length scale for $\bL^{\rm}_{0m0m'}$ involving solely $0$th-order DG coefficients, is noticeable longer than that for any higher-order DG 
coefficients. 

\begin{figure}[ht!]
\centering 
\includegraphics[width=\textwidth,height=.35\textheight,keepaspectratio=true]{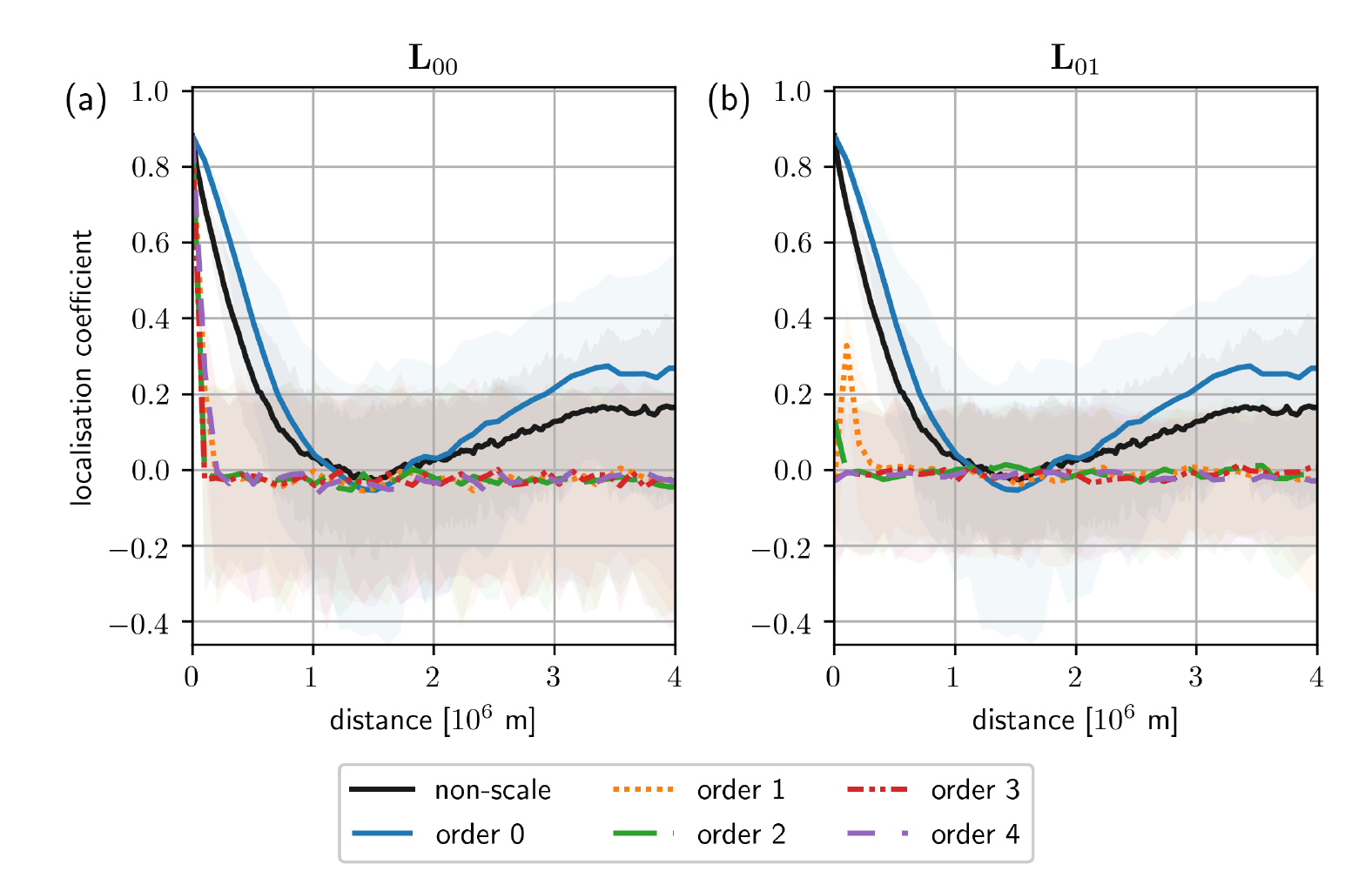}
\caption{As \fig{\protect\ref{fig:locCoef96}} but now using ensembles of $N=16$ members.  {\label{fig:locCoef16}}}
\end{figure}

The same localisation factors but now for an ensemble of $N=16$ members are shown in \fig{\ref{fig:locCoef16}}. Qualitatively the dependence of the localisation factors on distance is comparable with that in \fig{\ref{fig:locCoef96}}, but with lower values.  Furthermore, \fig{\ref{fig:locCoef16}} and especially \fig{\ref{fig:locCoef96}} show that the assumption that the covariance is (block-)diagonal in wavelet space as is made in e.g. \citet{deckmyn_wavelet_2005-1,pannekoucke_filtering_2007,chabot_diagnosis_2017} cannot be generalised when the Legendre polynomials are used as wavelets: \fig{\ref{fig:locCoef96}}b and \fig{\ref{fig:locCoef16}}b testify that over short distances cross-order localisation factors can take on values that are significantly different from zero. 

\subsection{Resulting Covariances}
{\label{sec:locCovariances}}

In \sect{\ref{sec:locCoefficients}} it was shown that scale-dependent localisation renders localistion factors that behave differently than those obtained by non-scale dependent localisation. In this section, we investigate whether this results in visibly different ensemble covariances and if these covariances are a better approximations of the true covariance than those obtained with non-scale dependent localisation. 

\begin{figure}[p]
    \centering
    \includegraphics[width=\textwidth,height=0.9\textheight,
    keepaspectratio]{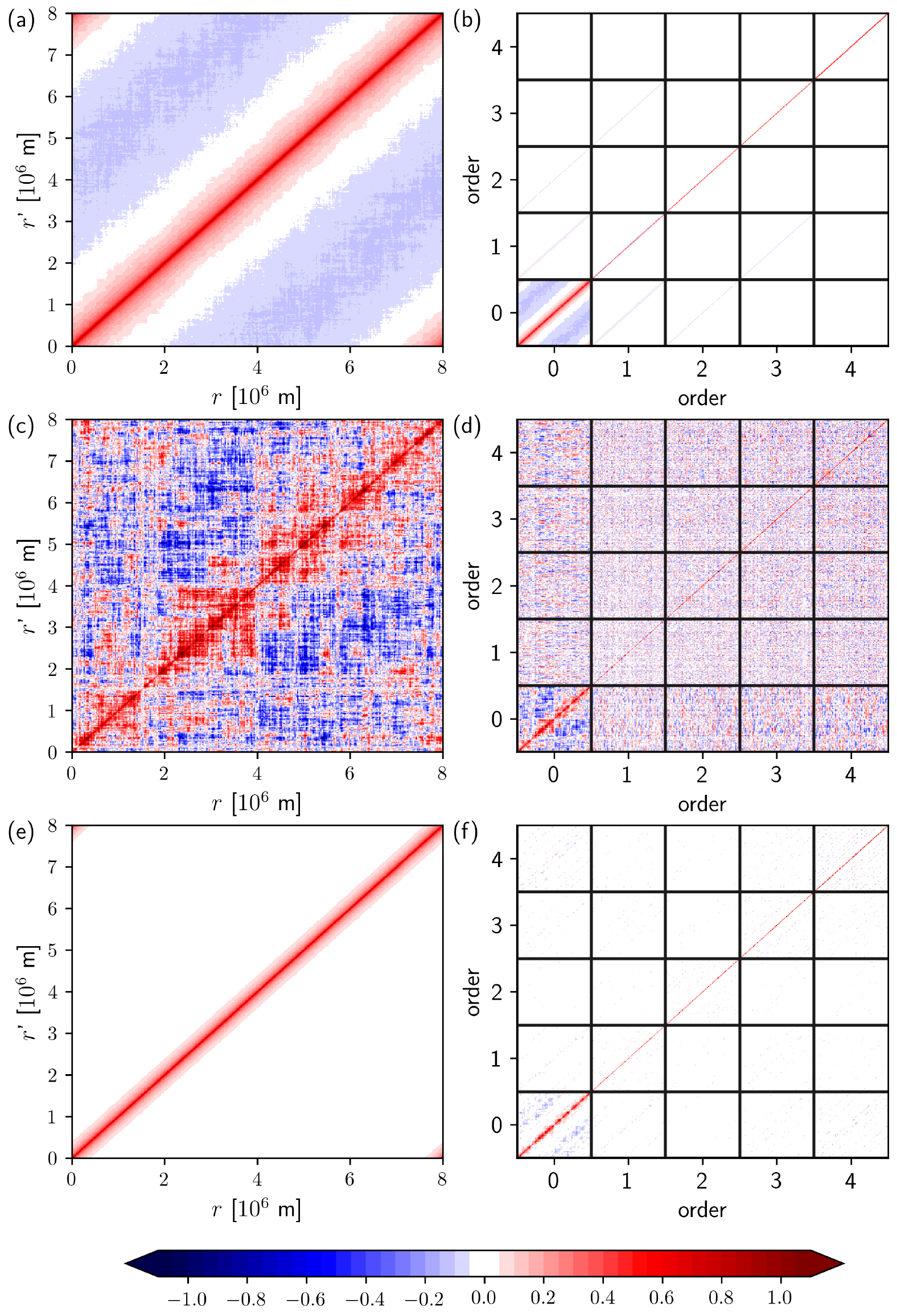}
    \caption{The (left column) nodal ensemble covariances and (right column) DG ensemble covariances (a,b) from a $10,000$-member, (c,d) 16-member ensemble without localisation and (e,f) 16-member ensemble with localisation. \label{fig:covariances}}
\end{figure}

The $10,000$ ensemble members are converted to $10,000$ \textit{nodal states} and $10,000$ \textit{DG states}. The covariance of the $10,000$ \textit{nodal states} is shown in \fig{\ref{fig:covariances}}a. Positive covariance up to $1.03$ can be found in a band near the diagonal. Because of periodicity, this band is repeated near $(0, 8\cdot 10^{6} \,\unit{m})$ and $(8\cdot 10^{6} \,\unit{m})$. Between these bands, covariances are negative with a low of $-0.15$.The equivalent covariance of \textit{DG states} is shown in \fig{\ref{fig:covariances}}b. Each box in \fig{\ref{fig:covariances}}b is a spatial covariance, similar to the one in \fig{\ref{fig:covariances}}a, but only involving DG coefficients of specific orders. For example, the 2nd box (from left) on the bottom row shows the covariance $\frac{1}{N-1}\sum_{n=1}^{N} \bx^{(n)}_{0m}\bx^{(n)}_{1m'}$, which is the cross-covariance between the coefficients of polynomials of order $0$ and $1$ at $m$th and $m'$th grid cell respectively. We will refer to the covariance as shown in this block as the $(0,1)$-DG covariance. The $(0,0)$-DG covariance (bottom, left box in \fig{\ref{fig:covariances}}b) mimics the nodal covariance. For the $(i,j)$-DG covariances with $i>0$ and/or $j>0$, non-zero covariances are located in a small band near the diagonal with some blocks exhibiting no covariance at all. So, when a Legendre basis is used, DG covariances exhibit distinct dependence on the order of the DG coefficient. 

Not only the covariance itself, but the sampling error too depends on the order of the DG coefficients. The nodal covariance for the $16$-member ensemble is shown in \fig{\ref{fig:covariances}}c. As a consequence of the sampling errors the range in values is larger ($[-1.23,2.09]$) and the band structure is less discernible: we see how spurious (unrealistic) correlations appear. From the $16$-member DG case in \fig{\ref{fig:covariances}}d we see again that the $(0,0)$-DG covariance mimics the nodal covariance, but the sampling noise in the other blocks of fig{\ref{fig:covariances}}d exhibits less spatial structure than the error in the $(0,0)$-DG covariance: the higher-order DG covariances take the form of unstructured 
noise or unstructured noise plus positive covariance along the diagonal. I.e. off-diagonal the signal-to-noise ratio is smaller than one in these covariances. This rapid drop in signal-to-noise with spatial distance between points explains why the localisation factors for the higher orders in \fig{\ref{fig:locCoef16}}b
and \fig{\ref{fig:locCoef96}}b drop rapidly with distance. The noise in \fig{\ref{fig:covariances}}d,e is efficiently removed by the localisation (see  \fig{\ref{fig:covariances}}e and \fig{\ref{fig:covariances}}f). However, comparison of \fig{\ref{fig:covariances}}a and \fig{\ref{fig:covariances}}b with \fig{\ref{fig:covariances}}e and \fig{\ref{fig:covariances}}f respectively shows that it also removes the genuine negative covariances and reduces the width of the diagonal bands. This overzealous removal is not as much of a problem for the $(i,j)$-DG covariances with $i+j>0$ as these covariances are truly near-diagonal. 

In the aforementioned we have established that both the covariance and the sampling errors in the covariance differ depending on the order of the DG coefficients. Now we want to determine whether accounting for those differences in the localisation procedure yields better estimates of the true covariance. To enable side-by-side comparison either the nodal covariance must be converted into its representation in DG space or the DG covariance must be converted into a covariance in nodal space. A linear transformation that converts a \textit{DG state} into a \textit{nodal state} can be formulated as 
\begin{equation}
    \coordTrans =  \sum_{j=0}^{J-1}   \sum_{m=1}^{M}
    \contravec_{j} \big( \bigotimes_{l'=0}^{L-1} 
    \phi_{l'm}(r_{j}) \delta_{l l'}  \covec_{m} \big) ,
    \label{eq:coordTrans}
\end{equation}
with $r_{j} = \frac{\ell}{J}(j-1)$ the position of the $j$th nodal point. As it is a linear transformation the nodal representation of the localised DG covariance is given as 
\begin{equation}
    \coordTrans (\bB^{\rm{DG}} \circ \bL^{\rm{DG}}) \coordTrans\T .
    \label{eq:coordTransDg2Nodal}
\end{equation}
Since in our experiments the dimension of the nodal space is equal to that of the DG space, \eq{\ref{eq:coordTrans}} can be inverted to get the localised nodal covariance in the DG space as 
\begin{equation}
    \coordTrans^{-1} (\bB^{\rm{NM}} \circ \bL^{\rm{NM}}) \coordTrans^{-\rm{T}}.
    \label{eq:coordTransNodal2Dg}
\end{equation}

\begin{figure}[ht!]
    \centering
    \includegraphics[width=\textwidth,height=0.6\textheight,keepaspectratio=true]{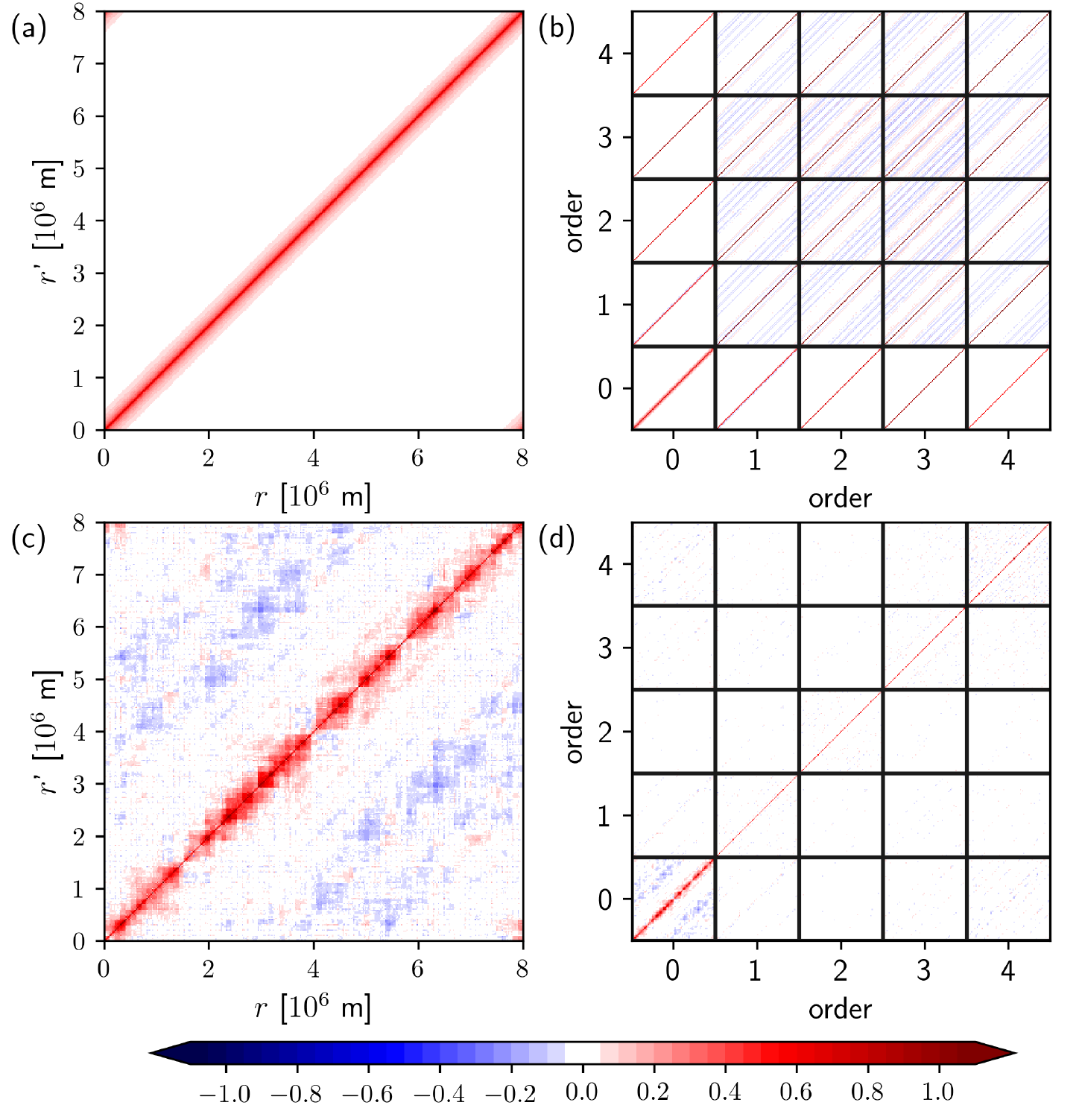}
    \caption{(a) 16-member ensemble covariance localised in nodal space and (b) its transformation into DG space. (d) The same 16-members but now localised in DG space using scale-dependent localisation together with (c) its transformation to nodal space.
    \label{fig:covarianceTrans}}
\end{figure}

Similar to \fig{\ref{fig:covariances}e}, the ensemble covariance of $16$ \textit{nodal states} after non-scale dependent localisation is shown in \fig{\ref{fig:covarianceTrans}a}. Next to it in \fig{\ref{fig:covarianceTrans}b} is its representation in DG space obtained by applying \eq{\ref{eq:coordTransNodal2Dg}} to \fig{\ref{fig:covarianceTrans}a}. If we compare the figure with the true DG covariance in \fig{\ref{fig:covariances}b} we see that for the $(i,j)$-DG covariances with $i>0$ and/or $j>0$ spurious localisation appear as bands parallel to the diagonal. This indicates that a strategy in which \textit{DG states} are mapped to and form \textit{nodal states} and in which localisation and DA is carried out in nodal space is troublesome: for the high-order DG coefficients the effective background error covariances obtained in this way do not match the covariance from the ensemble of {\it DG states} (\fig{\ref{fig:covariances}b}).
On the other hand, if we start of with the {\it DG states}, localise in DG space (\fig{\ref{fig:covarianceTrans}}d) and map this to nodal space using \eq{\ref{eq:coordTransDg2Nodal}} we get a covariance (\fig{\ref{fig:covarianceTrans}}c) that, although not flawless, better captures width of the true covariance (\fig{\ref{fig:covariances}}a) around the diagonal as well as the occurrence of negative
covariances than the covariance obtained using non-scale dependent localisation  (\fig{\ref{fig:covariances}}e). 

\begin{figure}[ht!]
    \centering
    \includegraphics[width=\textwidth,height=.35\textheight,keepaspectratio=true]{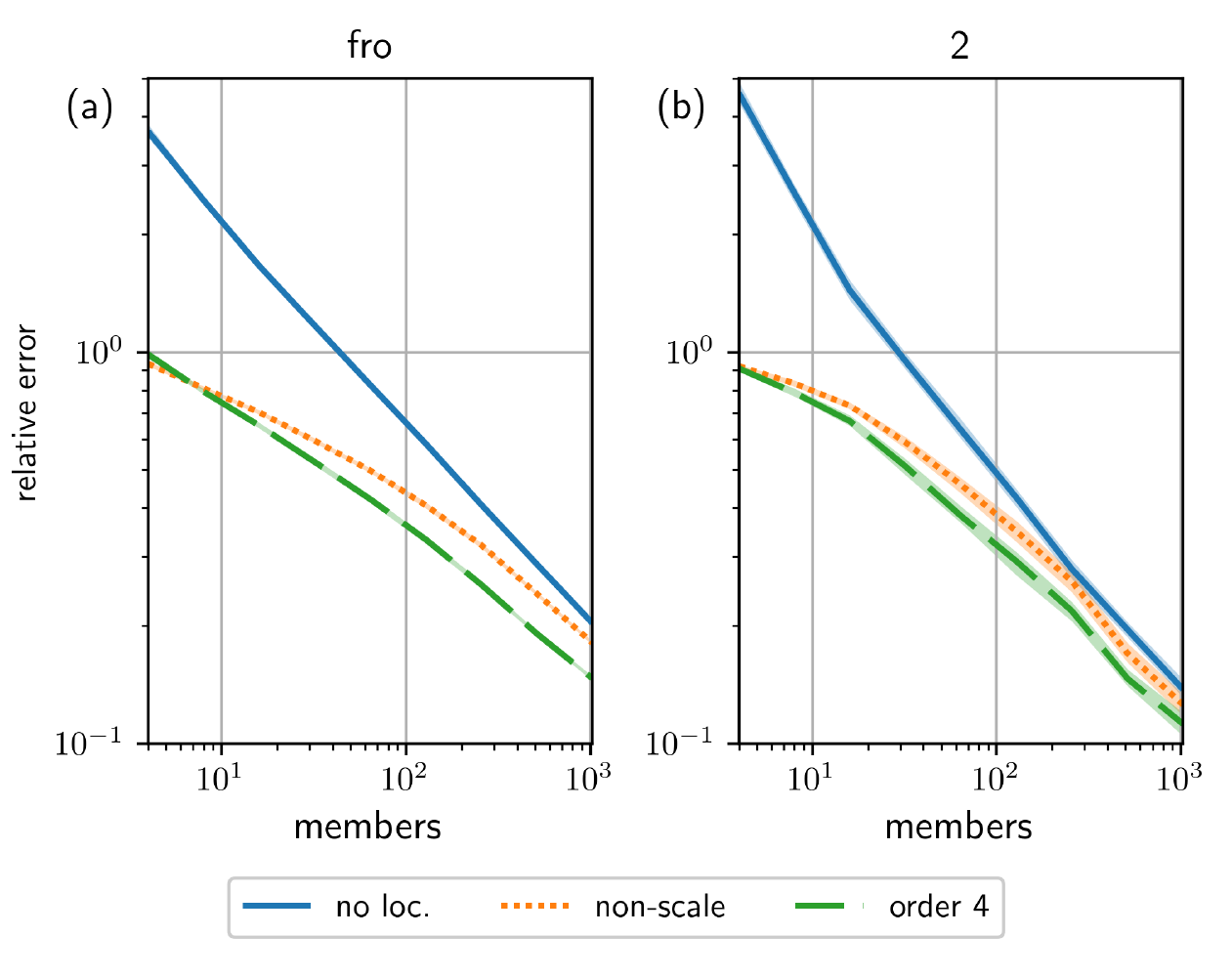}
    \caption{(a) Relative Frobenius and (b) operator/spectral error between the true ensemble covariance in nodal space as function of the number of ensemble members $N$ for covariance (blue) without localisation, (orange) with non-scale-dependent localisation, (green) with scale-dependent localisation. The $90\%$-confidence intervals in the relative error is shown as opaque shading.  {\label{fig:covarianceError}}}
\end{figure}    

To quantify the differences visible in \fig{\ref{fig:covarianceTrans}}, the relative error between the true nodal covariance and the covariance of \textit{nodal states} without localisation and  {\it nodal states} with non-scale-dependent localisation is shown in \fig{\ref{fig:covarianceError}} for different ensemble sizes. Also shown in the figure is the covariance created from {\it DG states} using scale-dependent localisation after mapping it to the nodal space. The error norm in the nodal space is calculated using two different methods: the Frobenius norm $\sqrt{\sum_{j=0}^{J-1} \sum_{j'=0}^{J-1} (\Delta \bB_{ll'})^2 }$ in \fig{\ref{fig:covarianceError}}a, and the operator/spectral norm which is the largest absolute eigenvalue of the error $\Delta \bB$ in \fig{\ref{fig:covarianceError}}b. The errors are normalised by the Frobenius or spectral norm of the true covariance respectively to obtain the relative error. The computation of the norms is repeated $100$ times with different ensembles and the mean of the relative norms and its $90\%$-confidence interval is shown in \fig{\ref{fig:covarianceError}}. Both non-scale and scale-dependent localisation reduce the relative error strongly compared to no-localisation with the reduction being more profound for the smaller ensemble sizes. The scale-dependent localisation produces relative errors that for $N\geq8$ ensemble size are slightly, but significantly better at the $90\%$-confidence level, than the covariances using non-scale-dependent localisations. In summary, non-scale dependent localisation applies too much localisation to the largest scales in the errors, resulting in the removal of the negative, off-diagonal covariances, and too little to the smallest scales leaving spurious off-diagonal covariances for the higher-order DG coefficients. Scale-dependent localisation addresses these issues a renders an overall more accurate representation of the background error covariance matrix. 

\section{Discussion and Conclusions}
{\label{sec:conclusion}}

Discontinuous Galerkin models resolve the solution of a system of partial differential equations as a linear combination of basis functions. We have shown in this study that if Legendre polynomials are used as basis functions, these functions constitute a basis of wavelets: each polynomial in each grid cell represents portions of the solution in specific portions of the physical space and of the spectral domain. Furthermore, the polynomials resolve the solution within grid cells, not solely on a finite-number of points as is the case in finite-difference or finite-volume methods. Both these characteristics can, in principle, be exploited by DA. This paper is a first exploration along this line. Here we have used a highly-idealised DA setup to investigate if, and under what conditions, these theoretically benefits can be realised. We tackled three aspects: the dependence of the DA performance on the observation density, the ability of DA-DG combination to improve estimates of the derivatives and the DG-based scale-dependent localisation. 

Dependency of RMSE ratios on the observation density show that DA in DG models benefits from increasing observation density and consequently that less data thinning/superobbing should be applied when deploying DA in DG models. In agreement with the data-thinning finding in \citet{liu_interaction_2002}, we see that the RMSE reduction rate levels off in all models, but in the DG models this happens at higher densities than in the nodal model. Based on our results we would suggest an observation density of 5 observations per grid cell to be a good compromise between achieving maximal error reduction and computational requirements. Further reduction of RMSE can be achieved by increasing the order of the DG scheme. The magnitude of the latter improvement, however, depends on the scales contained in the background errors. If small-scale errors dominate, increasing DG order results in a monotomic reduction of RMSE. On the other hands, if large-scale errors dominate, increasing DG order beyond $2$ yields no additional benefit. In both cases, most of the benefit is attained already for orders as low as $2$: an order that is achievable in operational models and will be used in \nextsimdg. Comparing the
RMSE analysis/background-ratio DG-order-by-order we found that the ratio is lower (better DA performance) when large scales dominate the background error. This is in agreement with the findings of \citet{fowler_interaction_2018-1} who showed that DA performs best when observations and background are accurate at different scales. As 
we have used white noise (thus small scale) to simulate observational errors, we get the greatest reduction in RMSE ratio if the large scales dominate the error background errors. 

This study could not find any indication that DG models can help DA with reducing errors in the derivatives of the field. When small-scale errors are dominant in the background errors of higher-order DG models, the improvements in DA performance actually come at the cost of deteriorating the estimates of the higher-order derivatives. This suggests that if accurate estimates for the derivatives are required then methods that include such derivatives in the cost-function are needed. One way to achieve this would be to assimilate the derivatives or their approximations as observations. A successful example of such an approach can be found in \citet{bedard_practical_2020} in which not only a thinned set of observations is assimilated, but also the differences between observation. The latter is found to reduce analysis errors for intermediate wavelengths. 

When the Legendre basis is used, it was found that the functional dependence of the optimal localisation coefficients depends strongly on the order of the DG coefficient and that consequently tuning localisation on the order of the DG coefficients is appropriate. When this is done, it produces covariances that are closer to the true background error covariance and are apter in retaining weak long-distance covariances. One of the major benefits of wavelet-based localisation already pointed out in previous works \citep{deckmyn_wavelet_2005-1,pannekoucke_filtering_2007,varella_diagnostic_2011} is its ability to represent correlations with length scales that vary in space. This aspect has not been covered in this study as we assumed our background error statistics to be homogeneous in space. Because of the similarities with the schemes used in aforementioned studies, the DG-based localisation scheme in this study should be able to deal with varying correlation scales too. However, it is important to note that when Legendre polynomials are used as basis functions, all scales larger than the grid cell width are grouped together. I.e. if the grid cell width is much smaller than all scales of interest, DG-based localisation is expected to tend to the same result as localisation in the nodal space. Another well-known issue with the optimal localisation scheme, and one that has not been addressed in this paper, is that localisation coefficient matrices produced by optimal localisation are not bounded to be positive definite. One proposed way to overcome this problem is to fit positive-definite kernels to $\bL$ \citep{michel_objective_2016}. Such a approach could also directly be applied to the DG-based scale-dependent localisation. Alternatively, positive-definiteness can be achieved by actively forcing the Fourier coefficients of the distance-dependent localisation function to be positive \citep{bochner_monotone_1933}. 

The work in this study was highly idealised. Nevertheless, its findings should be readily extendable to 2D and 3D models provided that they are defined on tensor product grids, e.g. cartesian grids. In that case, tensor products of the Legendre polynomials can be used as basis functions and the results in this paper are applicable in each direction separately. When using different grid topologies or basis functions, DA in higher-order DG models is still expected to produce greater error reduction as these models resolve the solution on a subgrid level. However, scale-dependent localisation might not be feasible as other basis functions do not necessarily constitute a wavelet basis. 

\section{Acknowledgements}\label{acknowledgements}

The authors acknowledge the support of the Scale-Aware Sea Ice Project (SASIP) funded by Schmidt Futures – a philanthropic initiative that seeks to improve societal outcomes through the development of emerging science and technologies. The authors would like to thank Piotr Minakowski and Jan Hesthaven for the their aid with understanding DG methods and V\`{e}ronique Dansereau for her guidance on sea ice modelling. 

\section{Conflict of interest}

Authors have no conflict of interest to declare. 

\bibliographystyle{agu}
\bibliography{Pasmans2023}

\newpage
\appendix

\section{Symbols}
\label{app:symbols}

Overview with mathematical symbols used in this paper together with their meaning. 

\begin{table}[h!]
    \centering
    
    \begin{tabulary}{\textwidth}{LLL}
    \hline 
    Symbol & Equation & Name \\
    \hline 
    $\bB$ && Background error covariance \\

    $\bBdg_{lm'l'm}$ && Background error covariance between coefficients $\bxdg_{lm}$ and $\bxdg_{l'm'}$ \\

    $\bandpass(x)(r,t)$ & $(x * \phi_{l})(r,t)$
    & Legendre band-pass filter. \\
    
    $D_{m}$ && Grid cell domain for the $m$th grid cell. \\
    
    $\Dref$ && Reference cell domain. \\
    
    $J$ && Number of non-zero Fourier coefficients.\\
    
    $L$ && Number of DG basis functions. \\
    
    $\bL^{\rm{DG}}$ && Localisation operator for covariance in DG space. \\

    $\bL^{\rm{NM}}$ && Localisation operator for covariance in nodal space. \\

    $\bL^{\rm{DG}}_{ll'}$ && Localisation operator for covariance between DG coefficients associated with Legendre polynomial $l$ and Legendre polynomial $l'$. \\
    
    $\bL^{\rm{DG}}_{lml'm}$ & & Localisation factor to be applied to covariance element $\bB^{\rm{DG}}_{lml'm'}$ \\
    
    $M$ && Number of grid cells. \\
    
    $N$ && Number of ensemble members. \\
    
    $\projection_{l}$ & $\sum_{m'=1}^{M} \contravec_{m'} \otimes \Big( \bigotimes_{l'=0}^{L-1}  \delta_{ll'} \covec_{m} \Big)$ & Projection that selects those DG coefficients associated with basis functions of order $l$. \\

    $\br$ && Position in the model domain. \\

    $\spectrum(x)$  & $\spectrum(x)(\kappa)=|\fourier(x)(\kappa)|^2$. & Power spectrum of the function $x$. \\

    $t$ && Time. \\
    
    $\trans_{m}$ & $\trans_{m}(\contravec_{m'})=\contravec_{m'- m - M \lfloor \frac{m'-m}{M} \rfloor }$ &
    Translation/shift of vector with $m$ on a grid with period $M$. \\

    $\xtrue$ && Exact solution to the system of partial differential equations. \\

    $x(\br,t)$ & $\sum_{l=0}^{L-1} \sum_{m=1}^{M} \bx_{lm}(t) \phi_{lm}(\br)$ & DG approximation to the solution of the system of partial differential equations. \\

    $\bxdg(t)$ &  $\bigotimes_{l=0}^{L-1} \sum_{m=1}^{M} \bx_{lm} \contravec_{m}$ 
    & Tensor containing the coefficients in front of the basis functions (DG coefficients). \\

    $\bxp(t)$ & $\sum_{m=1}^{M} \bxp_{km} \contravec_{m}$ 
    & Tensor containing the values of the model fields on the grid nodes. \\

    $\kappa$ && Wavenumber. \\
    
    $\refBase_{l}$ && The $l$th basis function defined on the reference domain $\Dref$. \\
    
    $\phi_{lm}$ && The $l$th basis function with support in the $m$th grid cell. \\
    
    $\Psi_{m}$ && Coordinate transformation from $\Dref$ to $D_{m}$. \\
    
    \hline
    \end{tabulary}
    
    \caption{Symbols used with their definitions and descriptions. {\label{tab:symbols}}}
    
\end{table}
\newpage

\section{Field Generation}
\label{app:generation}

Assume that the power spectrum density $\widehat{\spectrum}(x)$ of the field $x$ is given. 
Then a random field on a periodic domain of length $\ell$ is created as 
\begin{equation}
    x(r) = \Re\Big[ \sum_{j=0}^{\lfloor \frac{J}{2} \rfloor} 
    (\epsilon_{1,j}+i \epsilon_{2,j}) \sqrt{\widehat{\spectrum}(|\kappa_{j}|)} e^{i \kappa_{j} r}
    \Big]
\end{equation}
with $\Re[\cdot]$ the real part of $\cdot$, $\epsilon_{1,j}$, $\epsilon_{2,j}$ drawn from a normal distribution and $\kappa_{j} =  j \frac{2 \pi}{\ell}$ the wavenumber. Then 
\begin{equation}
    \fourier(x)(\kappa) = \frac{1}{2}\sum_{j=-\lfloor \frac{J}{2} \rfloor}^{\lfloor \frac{J}{2} \rfloor} (1+\delta_{j0})
    (\epsilon_{1,j}+i \sign(\kappa_{j}) \epsilon_{2,j}) \sqrt{\widehat{\spectrum}(|\kappa_{j}|)} \delta(\kappa-\kappa_{j})
\end{equation}
and 
\begin{equation}
\spectrum(x)(\kappa) = \expectation{\fourier(x)(\kappa)^{\dagger} \fourier(x)(\kappa)} =
\frac{1}{2}\sum_{j=-\lfloor \frac{J}{2} \rfloor}^{\lfloor \frac{J}{2} \rfloor} (1+\delta_{j0})
\widehat{\spectrum}(|\kappa_{j}|)\delta(\kappa-\kappa_{j})
\end{equation}
with $\cdot^\dagger$ denoting the complex conjugate. Then in its one-sided form 
$\spectrum(x)(|\kappa_{j}|) =  \widehat{\spectrum}(|\kappa_{j}|)$ as desired. 

After generation the signal can be converted into Legendre coefficients by exploiting the orthogonality condition of the Legendre polynomials $\refBase_{l}$
\begin{eqnarray}
    \bx_{lm} &=&  \frac{2l+1}{2} \int_{D_{m}} \phi_{lm}(r) x(r)\, dr 
    = \frac{2l+1}{2} \int_{m \dr}^{m \dr + \dr} \legendre{l}(\frac{2r-2 m \dr}{\dr}-1)x(r) \, \mathrm{d}r \nonumber \\
    &=& \frac{2l+1}{2} \Re\Big[ \sum_{j=1}^{\lfloor \frac{J}{2} \rfloor} 
    (\epsilon_{1,j}+i \epsilon_{2,j}) \sqrt{\widehat{\spectrum}(|\kappa_{j}|)} \int_{m \dr}^{m \dr + \dr} \legendre{l}(\frac{2r-2 m \dr}{\dr}-1) e^{i \kappa_{j} r} \, \mathrm{d}r
    \Big]  \nonumber \\
    &+& \frac{1}{2}\sqrt{\widehat{\spectrum}(|\kappa_{0}|)} \epsilon_{1,0} \int_{m \dr}^{m \dr + \dr}  \legendre{l}(\frac{2r-2 m \dr}{\dr}-1)  \,\mathrm{d}r
    \label{eq:legendre_projection}
\end{eqnarray}
with $\dr=\frac{\ell}{N}$ the grid cell width, $\frac{2}{2l+1}$ a normalisation factor. As the functions $\refBase_{l}$ are polynomials, the integrals in \eq{\ref{eq:legendre_projection}} can be carried out using integration-by-parts. 

\subsection{Computational Cost Scale-Dependent Localisation}
{\label{app:locCost}}

To apply covariance localisation in the D-E3DVar algorithm $\bB$ in \eq{\ref{eq:rcg}} needs to be replaced with $\bB \circ \bL$. As D-E3DVar is an iterative solver, it requires repeated application of $\bB \circ \bL$ to a vector of the form $\bH\T \bR^{-\frac{1}{2}} \chi$ with $\chi \in \R^{\Nobs}$. Here we provide an estimate of the algorithm complexity as measured by the number of floating-point operations, ``flop\footnote{Flop will be treated as a \textit{zero plural} to avoid confusion with flops as abbreviation for floating-point operations per second.}'' for short, necessary for the application of $\bB \circ \bL$ to a vector $\bH\T \bR^{-\frac{1}{2}} \chi$.  Flop necessary to calculate the latter vector are not included as they are necessary regardless whether localisation is used. The number of flop will be listed in big-O notation: $\bigo(N)$ means there is $c \in \R: \text{flop}<cN$. 

For given $l$ and $l'$,  $\bL^{\rm{DG}}_{lml'm'}$ can be calculated using the convolutions in \eq{\ref{eq:vv}} and \eq{\ref{eq:B2}}. Calculation of each term in the convolution in \eq{\ref{eq:vv}} requires $\bigo(MN)$ flop. The convolution can be carried out most efficiently using Fast-Fourier transform at the expense of $\bigo(M \log M)$ flops \citep{cooley_historical_1967}. For \eq{\ref{eq:B2}} $N$ convolutions need to be carried out at a total cost of $N M \log M$ flop and the results need to be multiplied and summed costing another $\bigo(MN)$ flop. The division, summation and scalar multiplication in \eq{\ref{eq:optimal_L}} require an additional $\bigo(M)$ flop bringing the total cost to calculate $\bL_{lml'm'}$ for all $m$, $m'$ combined to $\bigo(N M \log M)$. 

Each element of $(\bB^{\rm DG} \circ \bL^{\rm DG}) \chi$ can be written as
\begin{equation}
    \big( (\bB^{\rm DG} \circ \bL^{\rm DG})\chi \big)_{lm} =  \sum_{l'=0}^{L_k-1} \sum_{m'=1}^{M}
    \frac{1}{N-1} \sum_{n=1}^{N} \ba_{lm}^{(n)} \ba_{l'm'}^{(n)} \bL^{\rm DG}_{lml'm'} \chi_{l'm'} .
    \label{eq:BL}
\end{equation}
Based on the foregoing for each $l$, $\bigo(L N M \log M)$ flop are necessary for $\bL^{DG}_{lml'm'}$. For each $l$ and $m$ $\bigo(L M N)$ multiplications/additions appear in \eq{\ref{eq:BL}}. 
Thus, in total, the calculation of $(\bB^{DG} \circ \bL^{DG})\chi$ requires
$ \sum_{l=0}^{L-1} \bigo(L N M \log M) +M \bigo(L M N)=\bigo(L^2 N M \log M)+\bigo(L^2 N M^2)$ flop. 

The number of flop for non-scale dependent localisation on a grid with $M^{\rm NM}$ nodes, can be found by setting $L=1$ in the foregoing and is 
$\bigo(N M^{\rm NM} \log M^{\rm NM})+\bigo(N (M^{{\rm NM}})^{2})$. So, for equal number of grid cells, $M^{\rm NM}=M$, the scale-dependent localisation is $L^2$ times more expensive than the non-scale dependent localisation. If the dimension of the state is equal in both cases, i.e. $M^{\rm{NM}}=L M$, then the scale-dependent approach is 
$\frac{1+(M)^{-1}\log M }{1+(L M)^{-1}\log (L M)}<1+(M)^{-1}\log M \leq 1+e^{-1}$ more expensive. So, if DG and nodal space have similar dimensions, scale-dependent localisation can be up to $37\%$ more expensive than non-scale dependent localisation. 

\end{document}